\documentclass[prl,aps,preprintnumbers,twocolumn, showpacs, keywords, acknowledgement]{revtex4}

\usepackage{amsmath,amssymb,bm}
\usepackage{graphicx}
\usepackage{color}

\newcommand{\sign}{{\rm sign}}

\def\eqa{\begin{eqnarray}}
\def\eea{\end{eqnarray}}
\newcommand{\eq}{\begin{equation}}
\newcommand{\ee}{\end{equation}}

\begin{document}

\title{Realizing Majorana fermion modes in the Kitaev model 
}

\author{Lu Yang$^{1}$}
\author{Jia-Xing Zhang$^{1}$}
\author{Shuang Liang$^{3}$}
\author{Wei Chen$^{1,2}$} \email{chenweiphy@nju.edu.cn}
\author{Qiang-Hua Wang$^{1,2}$}
\affiliation{$^{1}$National Laboratory of Solid State Microstructures and School of Physics, Nanjing University, Nanjing, China}
\affiliation{$^2$Collaborative Innovation Center of Advanced Microstructures, Nanjing University, Nanjing, China}
\affiliation{$^3$Institute of Physics, Chinese Academy of Sciences, Beijing 100190, China}

%

%

\begin{abstract}
We study the possibility to realize Majorana zero mode that's robust and may be easily manipulated for braiding in quantum computing in the ground state of the Kitaev model  in this work. 
To achieve this we first apply a uniform $[111]$ magnetic field to the gapless Kitaev model and turn the Kitaev model to an effective $p+ip$ topological superconductor of spinons. We then study possible vortex binding in such system to a topologically trivial spot in the ground state. We consider two cases in the system: one is a vacancy and the other is a fully polarized spin. We show that in both cases, the system binds a vortex with the defect and a robust Majorana zero mode  in the ground state at a weak uniform $[111]$ magnetic field. The distribution and asymptotic behavior of these Majorana zero modes is studied. 
The Majorana zero modes in both cases decay exponentially in space, and are robust against local perturbations and other Majorana zero modes far away, which makes them promising candidate for braiding in topological quantum computing.
\end{abstract}

\pacs{75.10.Kt, 75.10.Ha}

%

\maketitle

\date{\today}

\section{I. Introduction}

The search for Majorana fermion (MF)  modes in condensed matter systems has intrigued great interest in recent years\cite{Read2000, Kitaev2001, Kouwenhoven2012, Yazdani2017, Kouwenhoven2019, Oreg2010, Sau2010, Sau2012, Fulga2013, Moore1991, Rice1995, Fu2008, Yazdani2013, Silaev2010}. One important reason is because a  pair of widely separated MF bound states is immune to local perturbations and  may be used for fault-tolerant quantum memory~\cite{Oreg2010}. Moreover, the non-Abelian statistics the MF states obey due to the degeneracy of such states may suggest them as components of a topological qubit and  be used in quantum information processing~\cite{Oreg2010, Nayak2008}.

Many approaches have been proposed to realize MFs in condensed matter systems, such as fractional quantum Hall states~\cite{Moore1991}, superfluids in $^3$He-B phase~\cite{Silaev2010}, semiconductors with strong spin-orbit interaction in both 2D and 1D~\cite{Oreg2010, Sau2010}, coupled quantum dots~\cite{Sau2012, Fulga2013}, array of magnetic atoms on the surface of superconductors~\cite{Yazdani2013}, and intrinsic topological superconductors~\cite{Read2000, Fu2008, Rice1995}. All these approaches involve superconductors in the system.

In this work, however, we discuss the realization of Majorana  modes in a  spin  system, the Kitaev model~\cite{Kitaev2006}. One possible realization of MF in the Kitaev model was on the edge of a chiral Kitaev spin liquid by applying a conical magnetic field on the gapless Kitaev model~\cite{Kitaev2006, Motome2019}. However, the edge Majorana zero mode is embedded in other edge modes with dispersion, which makes it easy to decay to other edge modes. Another proposal to achieve Majorana zero modes in Kitaev model is by introduction of  vacancy in the Kitaev model~\cite{Willans2010, Willans2011}. In Ref.~\cite{Willans2010, Willans2011}, it was shown that a single vacancy in the gapless Kitaev model results in a zero mode in the ground state. However, this Majorana zero mode is embedded in a continuum of spectrum and decays algebraically in space. For the reason, it interacts strongly with other vacancy induced zero modes  and disappears even if the other zero modes  are far away~\cite{Willans2011}.   These Majorana zero modes are then not robust enough for quantum computing.

In this work, we propose a realization of MF modes in the Kitaev model  based on the result that the Kitaev model in a weak conical magnetic field turns into an effective $p+ip$ superconductor of spinons~\cite{Kitaev2006}. And it's known that vortex in a p-wave superconductor may bind a Majorana zero mode that's robust against local perturbation~\cite{Read2000, Volovik1999, Xiang2007}. However, these vortex excitations are usually energy-costing~\cite{Xiang2007}. In this work, we discuss the possibility to create robust and easily manipulated Majorana zero mode in the ground state of the Kitaev model as an analog of the vortex bound Majorana zero mode in the p-wave superconductor. We consider two types of defects which may result in vortex binding (or $\pi$-flux-binding) in the ground state of the Kitaev model. One type of defect is the vacancy in the Kitaev model with a uniform $[111]$ magnetic field. We show that under a weak uniform $[111]$ magnetic field, a vacancy in the Kitaev model binds a  flux in the ground state which results in a  Majorana zero mode that decays exponentially in space. The second type of defect we study is a locally polarized spin achieved by a local magnetic field in the Kitaev model with a weak uniform $[111]$ magnetic field. We show that upon the full polarization of the local spin, a flux is  bound to the local spin plaquette in the ground state, which also results in a robust Majorana zero mode whose wavefunction  decays exponentially in space. The distribution and asymptotic behaviors of the Majorana zero modes in the above two cases are also studied in this work. In both cases, the Majorana zero modes are immune to local potential perturbations and other Majorana zero modes far away, and then may be used for braiding in topological quantum computing.

This paper is organized as follows: In Sec.II, we have a brief introduction of the model. In Sec.III, we study the vacancy induced Majorana zero mode in the Kitaev model with a weak uniform $[111]$ magnetic field, including the distribution and asymptotic behavior of the Majorana zero mode in both the continuum limit and lattice model, as well as the regime in which the flux binding to the vacancy takes place in the ground state. Sec. IV studies the Majorana zero mode induced by the polarization of a local  spin. We show the magnetization process of the local spin by mean field theory and the flux binding in the ground state upon polarization. The distribution of the Majorana zero mode after polarization is also studied in this section. At last, we  compare the Majorana zero mode bound to the vacancy with the edge Majorana modes in the system and summarize the main results in this work.

\section{II. Model and A brief review}

\begin{figure}
	\includegraphics[width=8cm]{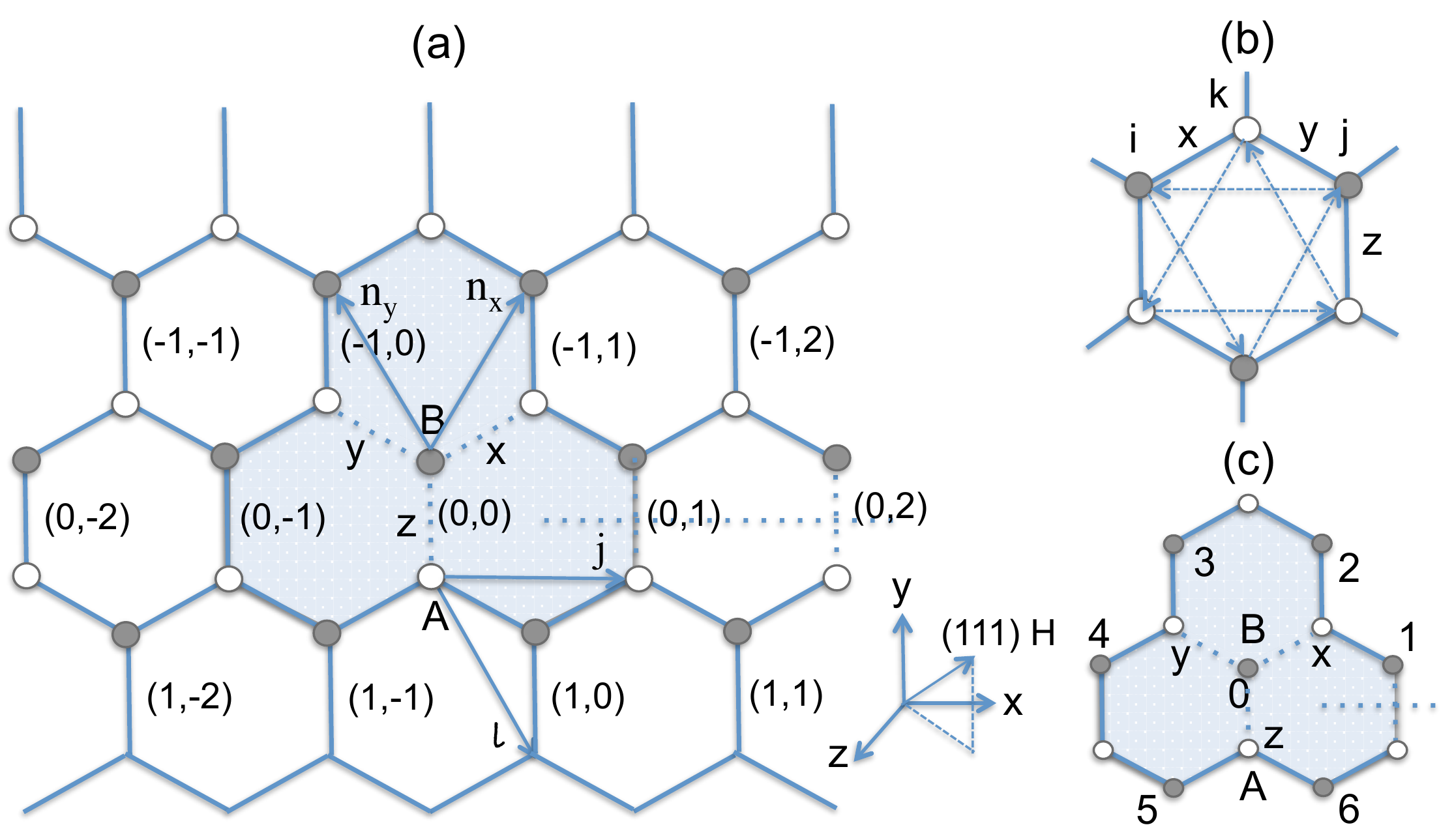}
		\caption{(a)The honeycomb lattice and the $x, y, z$ bond in the Kitaev model with a vacancy site at the B sublattice at $(0,0)$. Each unit cell contains a $z$ bond. The flipping of bond operator $u_{\langle ij\rangle_z}$ on the $z$ bonds crossing the half-infinite  dashed line produces a flux/vortex in the shaded area. (b) The site index of the third order perturbation Hamiltonian in Eq.(\ref{eq:perturbation}). The arrows indicate the hopping directions between the next nearest neighbors in the ground state.
		(c) Relabeling of site index with a local magnetic field applied on a B site labeled $0$ in Sec. IV.}\label{fig:lattice}
\end{figure}

The Kitaev model describes bond-dependent interaction of half spins on a honeycomb lattice with Hamiltonian~\cite{Kitaev2006}
\begin{eqnarray}
H_K=
-\sum_{\langle ij\rangle_\alpha} J_\alpha \sigma^\alpha_i \sigma^\alpha_j,
\end{eqnarray}
where $\alpha=x,y,z $ and $\langle ij\rangle_\alpha$ denotes two sites sharing an $\alpha$ bond.
The pure Kitaev model is solved by representing the half spin on each site with four Majorana fermions $\hat{c}_i, \hat{b}^x_i, \hat{b}^y_i, \hat{b}^z_i$ as $\sigma^\alpha_i=i\hat{c}_i \hat{b}^\alpha_i$~\cite{Kitaev2006} or a Jordan-Wigner transformation of half spins \cite{Xiang2007-2}. In this work, we adopt the Majorana representation. However, the Majorana representation enlarges  the physical Hilbert space of half spin by twice. For the reason, a local constraint $D_i=\hat{c}_i \hat{b}^x_i \hat{b}^y_i \hat{b}^z_i=1$ is imposed on each site
 to narrow down the states to the physical space. The pure Kitaev Hamiltonian then reads
\begin{equation}\label{eq:ground_state}
H_K=i\sum_{\langle ij\rangle_\alpha, \alpha} J_\alpha u_{\langle ij\rangle_\alpha} \hat{c}_i \hat{c}_j,
\end{equation}
where the bond operator $u_{\langle ij\rangle_\alpha}\equiv i \hat{b}^\alpha_i \hat{b}^\alpha_j$  is conserved with $u_{\langle ij\rangle_\alpha}=\pm1$. The ground state corresponds to the gauge invariant flux $W\equiv\prod_{\pi} u_{\langle ij\rangle_\alpha}$ defined on each hexagon $\pi$ to be $1$
and the spectrum of Eq.(\ref{eq:ground_state}) is $\epsilon_{\textbf{q}}=\pm |s(\textbf{q})|$, where $s_\textbf{q}=J_x e^{i \textbf{q} \cdot \textbf{n}_x}+J_y e^{i \textbf{q}\cdot  \textbf{n}_y}+J_z$~\cite{Kitaev2006} and $\textbf{n}_x, \textbf{n}_y$ are shown in Fig. \ref{fig:lattice}a.

To realize Majorana zero modes in the bulk of the Kitaev lattice in this work, we first apply a $[111]$ magnetic field to the Kitaev model, as shown in Fig.\ref{fig:lattice}a, which turns the Kitaev model to an effective $p_x+i p_y$  superconductor of spinons~\cite{Kitaev2006}. The Hamiltonian then becomes 
\begin{eqnarray}
H&= &H_K+H_h\nonumber\\
&=&-\sum_{\langle ij\rangle_\alpha} J_\alpha \sigma^\alpha_i \sigma^\alpha_j-h \sum_{i, \alpha}  \sigma^\alpha_i .
\end{eqnarray}

The uniform $[111]$ magnetic field breaks the conservation of the flux operator $W$ of each hexagon plaquette. However,  at weak $[111]$  magnetic field, the most important contribution of the  magnetic field comes from the third order perturbation theory~\cite{Kitaev2006}
\begin{equation}\label{eq:perturbation}
H_h=-t \sum_{\langle ijk \rangle}\sigma^x_i \sigma^y_j \sigma^z_k,
\end{equation}
where the configuration of the three neighboring sites $i, j, k$ are shown in Fig. \ref{fig:lattice}b and $t\sim h^3/J^2$.
In this work we  focus on this small uniform magnetic field regime so perturbation theory is valid and we  only keep the term in Eq.(\ref{eq:perturbation}) for the uniform magnetic field. In the Majorana representation, the Kitaev Hamiltonian including the perturbative $[111]$ magnetic field is then~\cite{Kitaev2006}
\begin{eqnarray}\label{eq:gap_Hamiltonian}
H&=&H_K+H_h \nonumber\\
& =& i\sum_{\langle ij\rangle_\alpha, \alpha} J_\alpha u_{{\langle ij\rangle}_\alpha} \hat{c}_i \hat{c}_j+t \sum_{\langle\langle ij \rangle\rangle} i \epsilon_{\alpha\beta\gamma} u_{{\langle ik\rangle}_\alpha} u_{{\langle kj\rangle}_\gamma}\hat{c}_i \hat{c}_j,\nonumber\\
\end{eqnarray}
where $\langle\langle ij \rangle\rangle$ represents next nearest neighbor $i$ and $j$ and $\alpha, \beta, \gamma=x, y, z$ represents the three bonds connected to site $k$.

With only the contribution from the third order perturbation of the $[111]$ magnetic field, the flux operator $W$ for each honeycomb plaquette is still conserved. In the ground state sector, $u_{{\langle ij\rangle}_\alpha}=1$ and the direction of the next nearest neighbor hopping is shown in Fig.\ref{fig:lattice}b. 
The ground state Hamiltonian is then 
\begin{eqnarray}\label{eq:GS_Hamiltonian}
H_0= i\sum_{\langle ij\rangle_\alpha, \alpha} J_\alpha  \hat{c}_i \hat{c}_j+i t \sum_{\langle\langle ij \rangle\rangle} i\hat{c}_i \hat{c}_j.
\end{eqnarray}
After Fourier transformation to the momentum space $\hat{c}_{\textbf{q},A}, \hat{c}_{\textbf{q}, B}$, where  $\hat{c}_{\textbf{q}, A/B}=\frac{1}{\sqrt{2N}}\sum_{\textbf r} e^{-i{\textbf q} \cdot \textbf{r}} \hat{c}_{{\textbf r}, A/B}=\hat{c}^\dag_{-\textbf{q}, A/B}$ becomes a complex fermion operator, the Hamiltonian Eq.(\ref{eq:gap_Hamiltonian}) becomes~\cite{Kitaev2006}
\begin{eqnarray}\label{eq:Fourier_Hamiltonian}
H=
\sum_\textbf{q} \hat{\psi}^\dag_\textbf{q}  \left( \begin{array}{cc}
   \Delta(\textbf{q})   &\  \ s(\textbf{q})\\ 
 \ s(\textbf{q})^* &  \ -\Delta(\textbf{q}) \\ 
  \end{array}\right)\hat{\psi}_\textbf{q},
\end{eqnarray}
where $\hat{\psi}_\textbf{q}=(\hat{c}_{\textbf{q},A}, i \hat{c}_{\textbf{q}, B})$, $s(\textbf{q})$ gives the ground state spectrum of the pure Kitaev model, and $\Delta(\textbf{q})=4t (\sin(\textbf{q}\cdot \textbf{n}_x)- \sin(\textbf{q} \cdot \textbf{n}_y)+\sin(\textbf{q} \cdot (\textbf{n}_y-\textbf{n}_x)))$. 
The spectrum with the perturbative magnetic field is then $E(\textbf{q})=\pm \sqrt{|s(\textbf{q})|^2+\Delta(\textbf{q})^2}$. For the isotropic gapless Kitaev model,  a gap opens up at $\pm\textbf{q}_D=( \frac{\sqrt{3}}{3}\pi, \pi)$ with the  value $\Delta=\Delta(\textbf{q}_D)=-\Delta(-\textbf{q}_D)=6\sqrt{3} t$~\cite{Kitaev2006}.

By expansion at $\pm\textbf{q}_D$ and keeping only the second order of $\delta \textbf{q}=\textbf{q}-\textbf{q}_D$, the Hamiltonian matrix of  Eq.(\ref{eq:Fourier_Hamiltonian}) for the isotropic  Kitaev model becomes
\begin{eqnarray}\label{eq:Nambu_Hamiltonian}
H_{\bf q}=
\left( \begin{array}{cc}
    \pm 6\sqrt{3} t(1-\frac{1}{4}(\delta q)^2)  & \sqrt{3}J \ (\mp\delta q_x-i  \ \delta q_y) \\ 
   \sqrt{3}J( \mp\delta q_x+ i \ \delta q_y)  &   \mp6\sqrt{3} t(1-\frac{1}{4}(\delta q)^2) \\ 
  \end{array}\right),
\end{eqnarray}
near the two Dirac point $\pm {\bf q}_D$. This Hamiltonian matrix is the same as that of a weak pairing $p_x+i p_y$ superconductor~\cite{Read2000}, though the basis here is not the Nambu spinor for superconductor but a pseudo-spinor composed of the two sublattice fields.

The above Hamiltonian can be diagonalized by the Bogoliubov transformation $\hat{\xi}_\textbf{q}=u_\textbf{q}\hat{c}_{\textbf{q},A}+i v_\textbf{q}\hat{c}_{\textbf{q}, B}$. The unit vector characterizing the direction of the pseudo-spinor $(u_\textbf{q}, v_\textbf{q})^T$ is $\textbf{n}_\textbf{q}=(\pm\sqrt{3}J\delta q_x, \sqrt{3}J\delta q_y, \Delta(\textbf{q}))/E_\textbf{q}$ near $q_D$ and $-q_D$ respectively. For $t>0$, $\textbf{n}_\textbf{q}$ maps the neighborhood of $\textbf{q}_D$ to the north hemisphere of the sphere of unit vector and the neighborhood of $-\textbf{q}_D$ to the south hemisphere.
As $\textbf{q}$ varies over the two-dimensional momentum space  (which may be mapped to an $S^2$ sphere surface~\cite{Read2000}), the unit vector $\textbf{n}_q$ sweeps through the whole unit sphere $S^2$, resulting in a non-trivial winding number $1$ or $-1$ depending on the sign of $t$ or the direction of the $[111]$ magnetic field~\cite{Kitaev2006}. Without loss of generality, we assume $t>0$ in the following.

Since the Kitaev model is gapped at finite $t$, the Chern number of the spinon bands is well defined and may be easily computed from the
eigenvectors of Eq.(\ref{eq:Nambu_Hamiltonian}), which is $\nu=\sign \Delta =\pm 1$~\cite{Kitaev2006}, the same as a weak-pairing p-wave superconductor.

The Majorana modes in the $p_x+i p_y$ superconductors have been  studied extensively in previous works~\cite{Read2000, Volovik1999, Ivanov2001}. There are  mainly two mechanisms to realize Majorana fermion modes in such systems~\cite{Read2000}. One is on the edge of the superconductor, where the Majorana fermion modes may have  both zero energy and finite energy. The other is by creating vortex in such system, since a vortex may bind a Majorana zero mode in a gapped topologically non-trivial system with odd Chern number~\cite{Kitaev2006}. 

In the following sections, we discuss the realization of the Majorana fermion modes in the Kitaev spin liquid as an analog of the $p_x+i p_y$ superconductors. We focus on the realization of the Majorana zero modes by creating vortex in the above discussed  Kitaev model in the $[111]$ magnetic field. However, the vortex are usually energy costing in such system~\cite{Xiang2007, Jiang2020}. Our main goal in this work is then to discuss the possibility to create vortex and Majorana modes  in the ground state of this system which are suitable for braiding in topological quantum computing. For the edge mode, we only point out that for the lattice Kitaev model in the weak $[111]$ magnetic field, only the edge mode with zero energy corresponds to a Majorana mode whereas the chiral Majorana edge modes with finite energy only exist in the continuum limit of the Kitaev model.

\section{III. Vacancy induced Majorana zero modes}

In this section we investigate the possibility to create Majorana Fermion modes in the Kitaev model in the ground state  by introducing vacancies in the Kitaev model. It has been shown in Ref.~\cite{Willans2010, Willans2011} that a vacancy in the pure Kitaev model may combine a flux in the ground state. However, a gapped pure Kitaev model has Chern number zero and there is no robust Majorana zero mode bound to the flux. For the gapless Kitaev model with a vacancy, a zero-energy resonance state was discovered~\cite{Willans2010, Willans2011}. However, this state is embedded in the bulk gapless spectrum and is easy to decay to the bulk modes. What's more, this mode decays as a power law in space and two such modes interact and split to non-zero modes however far away they are. For the reason, these zero-modes are not suitable for braiding in quantum computing. To solve the above problem, we apply a uniform $[111]$ magnetic field to the gapless Kitaev model at first and turn the Kitaev model to an effective gapped $p_x+i p_y$ superconductor. We then study the vacancy in such system and possible flux binding and Majorana modes  in the ground state in such system.

\subsection{3.1. MZM induced by   a $\pi$ flux bound to a vacancy  in the continuum limit of the Kitaev model  under the $[111]$ magnetic field }

As an analog of the  $p_x+i p_y$ superconductor, the Majorana zero mode bound to a vacancy with a $\pi$ flux in the Kitaev model under the $[111]$ magnetic field may be solved in the continuum limit of the model as shown in Ref.\cite{Otten2019}. We briefly recapture the calculation 
here and compare it with the results from the lattice model in the next subsection. Different from the $p$-wave superconductor, the Majorana mode in the Kitaev model in the continuum limit is a superposition of Fermionic modes in the two valleys instead of one.

A Majorana Fermion field in the Kitaev model in the $[111]$ magnetic field may be expressed as 
\begin{equation}\label{eq:Majorana_field}
c_{A/B}({\bf r})=e^{i {\bf q_D \cdot r}} \phi_{A/B}({\bf r})+e^{-i \bf{ q}_D \cdot {\bf r}} \phi_{A/B}^\dag({\bf r}),
\end{equation}
where $\phi_{A/B}$ is a canonical Fermion field describing the contribution from one Dirac point $D$ and $\phi_{A/B}^\dag$ the contribution from the opposite Dirac point. Since the MZM in the above Kitaev model is a superposition of all the fields $\phi_{A}({\bf r}), \phi_{B}({\bf r}), \phi^\dag_{A}({\bf r}), \phi^\dag_{B}({\bf r})$, it's convenient to adopt the four component basis $\Phi(r)=(\phi_A(r), i\phi_B(r), \phi_A^\dag,  i \phi_B^\dag(r))$.  Inserting Eq.(\ref{eq:Majorana_field}) to the ground state Hamiltonian of  Eq.(\ref{eq:gap_Hamiltonian}) and neglecting the fast oscillating terms with $e^{\pm 2i {\bf q_D r}}$,
 the Hamiltonian Eq.(\ref{eq:gap_Hamiltonian}) in the continuum approximation may be written as 
\begin{equation}\label{eq:continuum_Hamiltonian}
H=\int d^2 r \Phi^\dag(r) {\cal H}(r)\Phi(r), 
\end{equation}
where ${\cal H}$ is block diagonal in the particle-hole or valley degrees of freedom. The upper  block is ${\cal H}_D=\Delta \sigma_z+\sqrt{3}J (i\sigma_x \partial_x-i\sigma_y\partial_y)$ and the lower block is ${\cal H}_{-D}=-\Delta \sigma_z+\sqrt{3}J (-i\sigma_x \partial_x-i\sigma_y\partial_y)$ keeping only the linear order of momentum. The two blocks ${\cal H}_D$ and ${\cal H}_{-D}$ may be diagonalized separately and for the zero mode
\begin{eqnarray}
{\cal H}_D \Phi_D&=&0, \label{eq:block_hamiltonian_1}\\
{\cal H}_{-D} \Phi_{-D}&=&0 \label{eq:block_hamiltonian_2}
\end{eqnarray}
where $\Phi_D$ and $\Phi_{-D}$ are two component spinors with basis $(\phi_A(r), i\phi_B(r))^T$ and $ (\phi_A^\dag,  i \phi_B^\dag(r))^T$ respectively.

Assuming the zero mode  $\Phi({\bf r})=u\phi_A({\bf r})+iv\phi_B({\bf r})+\tilde{u}\phi^\dag_A({\bf r})+i\tilde{v}\phi^\dag_B({\bf r})$,  the coefficients $u$, $v$ and $\tilde{u}, \tilde{v}$ satisfy
\begin{eqnarray}
 \Delta(\textbf{r}) u +i \sqrt{3} J (\frac{\partial }{\partial x} + i \frac{\partial }{\partial y} ) v &=& 0, \label{eq:one}\\
 - \Delta(\textbf{r}
) v +i \sqrt{3} J (\frac{\partial }{\partial x} - i \frac{\partial }{\partial y} ) u &=&0, \label{eq:two}\\
-\Delta(\textbf{r}) \tilde{u} +i \sqrt{3} J (-\frac{\partial }{\partial x} + i \frac{\partial }{\partial y} ) \tilde{v} &=& 0, \label{eq:three}\\
 \Delta(\textbf{r}
) \tilde{v} +i \sqrt{3} J (-\frac{\partial }{\partial x} - i \frac{\partial }{\partial y} ) \tilde{u} &=&0. \label{eq:four}
\end{eqnarray}
From the above equations, the coefficients $u$ and $v$ satisfy $u=\pm v^*$, and $\tilde{u}$ and $\tilde{v}$ satisfy $\tilde{u}=\pm \tilde{v}^*$. Moreover, the solution of the coefficients satisfy $\tilde{u}=u^*$ and $\tilde{v}=-v^*$ or $\tilde{u}=-u^*$ and $\tilde{v}=v^*$ by comparing Eq.(\ref{eq:one}) and Eq.(\ref{eq:three}).
To get a Majorana zero mode $\Phi_M({\bf r})=u\phi_A({\bf r})+iv\phi_B({\bf r})+\tilde{u}\phi^\dag_A({\bf r})+i\tilde{v}\phi^\dag_B({\bf r})=e^{i\gamma} \Phi^\dag_M({\bf r})$ where $e^{i\gamma}$ is merely a phase factor, we choose $u=v^*, \tilde{u}=-\tilde{v}^*, u=\tilde{u}^*, v=-\tilde{v}^*$. With this choice, $\Phi_M({\bf r})=\Phi^\dag_M({\bf r})$.

We then solve the MZM with a $\pi$ flux bound to a vacancy. From the above relationship, we only need to solve $u$ and $v$. And $\tilde{u}$ and $\tilde{v}$ may be obtained from the solution of $u$ and $v$.
The equations for $u$ and $v$ are the same as that of the zero mode for a $p_x+i p_y$ superconductor~\cite{Read2000}.
Far away from the vacancy, $\Delta(\textbf{r})$ becomes  $\Delta=6\sqrt{3}t$ in the uniform system.
In the vacancy center, $\Delta\to 0$. For a circular vacancy, the above equation can be easily solved  in the polar coordinate with the origin at the center of the vacancy
\begin{eqnarray}\label{eq:vortex_zero_mode}
&&\sqrt{3}J i e^{i \theta} (\frac{\partial}{\partial r}+\frac{i}{r}\frac{\partial}{\partial \theta}) v=-\Delta u, \\
&&\sqrt{3}J i e^{- i \theta} (\frac{\partial}{\partial r}-\frac{i}{r}\frac{\partial}{\partial \theta})u=\Delta v.
\end{eqnarray}
When a vortex is threaded to the vacancy as shown in Fig.\ref{fig:lattice}a with a cut along $\theta=0$, the quasiparticle wavefunction   obeys anti-boundary condition on going around the vortex and vanishes at the  vortex core, i.e., $u(r, \theta+2\pi)=-u(r, \theta)$ and $u(r\to 0)=0$, and the same for $v$. The above equation has a solution satisfying $u=v*=\frac{f(r)}{r^{1/2}}e^{i(\theta/2-\pi/4)}$, where $f(r)=e^{-\int^r_{r_0} \Delta dr/\sqrt{3}J}\sim e^{-\Delta r/\sqrt{3}J}$ and $r_0$ is the radius of the vacancy for $\Delta/J>0$. 
For $\Delta/J<0$, the decaying solution is $u=-v^*=\frac{f(r)}{r^{1/2}}e^{i(\theta/2-\pi/4)}$, where $f(r)=e^{\int^r_{r_0} \Delta dr/\sqrt{3}J}\sim e^{\Delta r/\sqrt{3}J}$.
With the above choice for the MZM, we get $\tilde{u}(r)=u^*(r)=\frac{f(r)}{r^{1/2}}e^{-i(\theta/2-\pi/4)}$ and $\tilde{v}=-v^*=-\frac{f(r)}{r^{1/2}}e^{i(\theta/2-\pi/4)}$.
The amplitude of the MZM  $\Phi_M({\bf r})$ decays exponentially with distance to the vacancy center as $\sim \frac{1}{r^{1/2}} e^{-|\bar{\Delta}\ r/\sqrt{3}J|}$, where $r$ is the distance to the vortex center and  $\bar{\Delta}=\int_0^r \Delta(\rho) d\rho/r$. This asymptotic behavior here  is the same as the vortex induced zero mode in the $p_x+ip_y$ superconductor~\cite{Read2000}, in contrast to the statement in Ref.~\cite{Otten2019} that the zero modes in the two systems have different asymptotic behaviours.

\subsection{3.2. Majorana zero modes in the lattice Kitaev model with a vacancy}

The zero mode associated with a vacancy in the Kitaev model under the $[111]$ magnetic field can also be obtained from the lattice Kitaev model. In this subsection, we show the similarities as well as the differences between the results from the continuum model and the lattice model.

The eigenstates of the lattice Kitaev model with a weak uniform $[111]$ magnetic field and a vacancy can be obtained by directly solving the bilinear Hamiltonian Eq.(\ref{eq:gap_Hamiltonian}) with an empty site. For the eigenstate with a vacancy site but flux-free, we set $u_{\langle ij \rangle_\alpha}=1$ for all the bonds. For the case with a $\pi$ flux bound to the vacancy, we set the $z$ bond operators $u_{\langle ij \rangle_z}=-1$ along the half-infinite string in Fig.\ref{fig:lattice}a. In both cases, the Hamiltonian  Eq.(\ref{eq:gap_Hamiltonian}) becomes the following form 
\begin{equation}\label{eq:bilinear_hamiltonian}
H=i\sum_{l, m} A_{lm} \hat{c}_l \hat{c}_m,
\end{equation}
where $A$ is a real anti-symmetric matrix and $\hat{c}_l, \hat{c}_m$ are Majorana fermions. The $A$ matrix then satisfies the following condition
\begin{equation}
A^T=-A, \ A^*=A,\  (iA)^\dag=iA.
\end{equation}
Since $iA$ is Hermitian, it can be diagonalized by a unitary matrix $\hat{U}$, i.e., 
\begin{equation}
\ \ \ \ \ \hat{U}^\dag \hat{U}=\hat{U} \hat{U}^\dag=1, \ iA \hat{U}=E\hat{U},\ \hat{U}^\dag (iA) \hat{U}=\hat{E},
\end{equation}
where $\hat{E}$ is a diagonal and real matrix. For an eigenvector $\vec{U}$ with eigenenergy $\epsilon$, i.e., $i A \vec{U}=\epsilon \vec{U}$, one can get $iA \vec{U}^*=-\epsilon \vec{U}^*$ from the above conditions. To obtain a Majorana eigenvector, $\vec{U}$ must satisfy $\vec{U}=e^{i\gamma}\vec{U}^*$ where $\gamma$ is a real constant, so $\epsilon=0$, i.e., the eigenenergy of a Majorana eigenmode must be zero for the lattice Kitaev Hamiltonian Eq.(\ref{eq:gap_Hamiltonian}) with a vacancy~\cite{Santhosh2012}.

For the eigenstates with non-zero energy, the eigenvectors $\vec{U}$ and $\vec{U}^*$ with energy $\pm \epsilon$ appear in pairs. For odd dimension of the $A$ matrix, there is at least one zero energy eigenmode. For even dimension of $A$, the number of zero energy eigenmodes is even.

For a lattice with N unit cell and a single vacancy site, the $A$ matrix becomes a $(2N-1)\times (2N-1)$ matrix. For both the flux-free case  and the case with a $\pi$ flux threading through the vacancy plaquette, there is a Majorana zero mode. Whereas for the continuum model in the last section, the zero mode equation Eq.(\ref{eq:one})-(\ref{eq:four}) has no solution in the flux free case.
The zero mode in the lattice model in the flux free case is then due to the specific lattice configuration and is not robust. When the configuration of the vacancy changes, e.g., if the vacancy in the above model includes two neighboring sites, the Majorana zero mode disappears  in the flux free case, which is verified in our numerics.
 However, the zero mode bound to the $\pi$ flux is robust  since it's induced by a topological defect and does not depend on the lattice configuration of the vacancy. 
Moreover, the zero modes in the two cases have different asymptotic behaviours which we will study in details in the following.

In the following we focus on the case with a single vacancy site in the lattice Kitaev model for simplicity
and compare the Majorana zero modes in the flux free and flux threading case. 
We only need to consider the vacancy on one of the two sublattices. The  amplitudes of the zero mode for the vacancy on the other sublattice may be obtained by inversion symmetry~\cite{Willans2010,Willans2011, Santhosh2012}.  Without loss of generality, we assume a vacancy on the B site in the following. We label the unit cell of the honeycomb lattice as in Fig. \ref{fig:lattice}a. Each unit cell contains a $z$ bond. For the flux free case, the amplitudes of the vacancy induced zero mode satisfy the equations away from the vacancy as follows (shown in the appendix)
\begin{eqnarray}
&&J_x a_{l-1, j+1}+J_y a_{l-1, j} +J_z a_{l, j} -t(b_{l-1,j}-b_{l-1, j+1} \nonumber\\
&&\ \ \ +b_{l, j+1}-b_{l, j-1}+b_{l+1, j-1}-b_{l+1, j})=0, \label{eq:A_and_B} \\ 
&&J_x b_{l+1, j-1}+J_y b_{l+1, j} +J_z b_{l, j} -t(a_{l-1,j}-a_{l-1, j+1} \nonumber\\
&&\ \  \ +a_{l, j+1}-a_{l, j-1}+a_{l+1, j-1}-a_{l+1, j})=0, \label{eq:B_and_A}
\end{eqnarray}
where $a_{lj}$ and $b_{lj}$ is the amplitude of the zero mode on the A and B sublattice of the unit cell $(l,j)$ respectively, with the basis being the Majorana fermion  $\hat{c}$ on each site. The boundary condition near the vacancy is: (a) Eq.(\ref{eq:A_and_B}) does not exist at $(l, j)=(0, 0)$; (b)For other $l, j=-1, 0, 1$, Eq.(\ref{eq:A_and_B})  and Eq.(\ref{eq:B_and_A}) is satisfied with the amplitude $b_{0, 0}=0$.

At $t=0$ and flux free, this zero mode was solved  in Ref~\cite{Pereira2006, Willans2011, Santhosh2012}. It locates only on the opposite sublattice of the vacancy site since the A and B sublattice decouples in the zero mode in this case and the recursion factor of the amplitudes on the A and B sublattice is reciprocal, as shown in the appendix. For the gapless Kitaev model and flux free case, the amplitude of the zero mode on the A sublattice away from the vacancy on a B site has the asymptotic form as~\cite{Pereira2006, Santhosh2012}
\begin{equation}\label{eq:amplitude}
a(x, y)\sim {\rm Re} \{ \frac{e^{i q_c x /\sqrt{3}+2i(\pi+\theta^*)y/3}}{(\alpha -2i\beta)y/\sqrt{3}-i x }  \},
\end{equation}
where $\alpha=2J_x J_y \sin(q_c)/J^2_z, \ \beta=(J^2_x-J^2_y)/J^2_z, \ q_c=\arccos \frac{J^2_z-J^2_x-J^2_y}{2J_x J_y}, \ \theta^*=\arctan [\frac{J_x-J_y}{J_x+J_y}\tan(q_c/2)]$ are constants, and $x=\sqrt{3}(j+\frac{1}{2}l),  y=\frac{3}{2}l$. The amplitude decays as a power law $1/r$ with distance $r$ from the vacancy, and when there are two vacancies in the system, the two zero modes couples strongly with each other and split to two finite energy levels however far away they are~\cite{Willans2011}.

\begin{figure}
	\includegraphics[width=7.5cm]{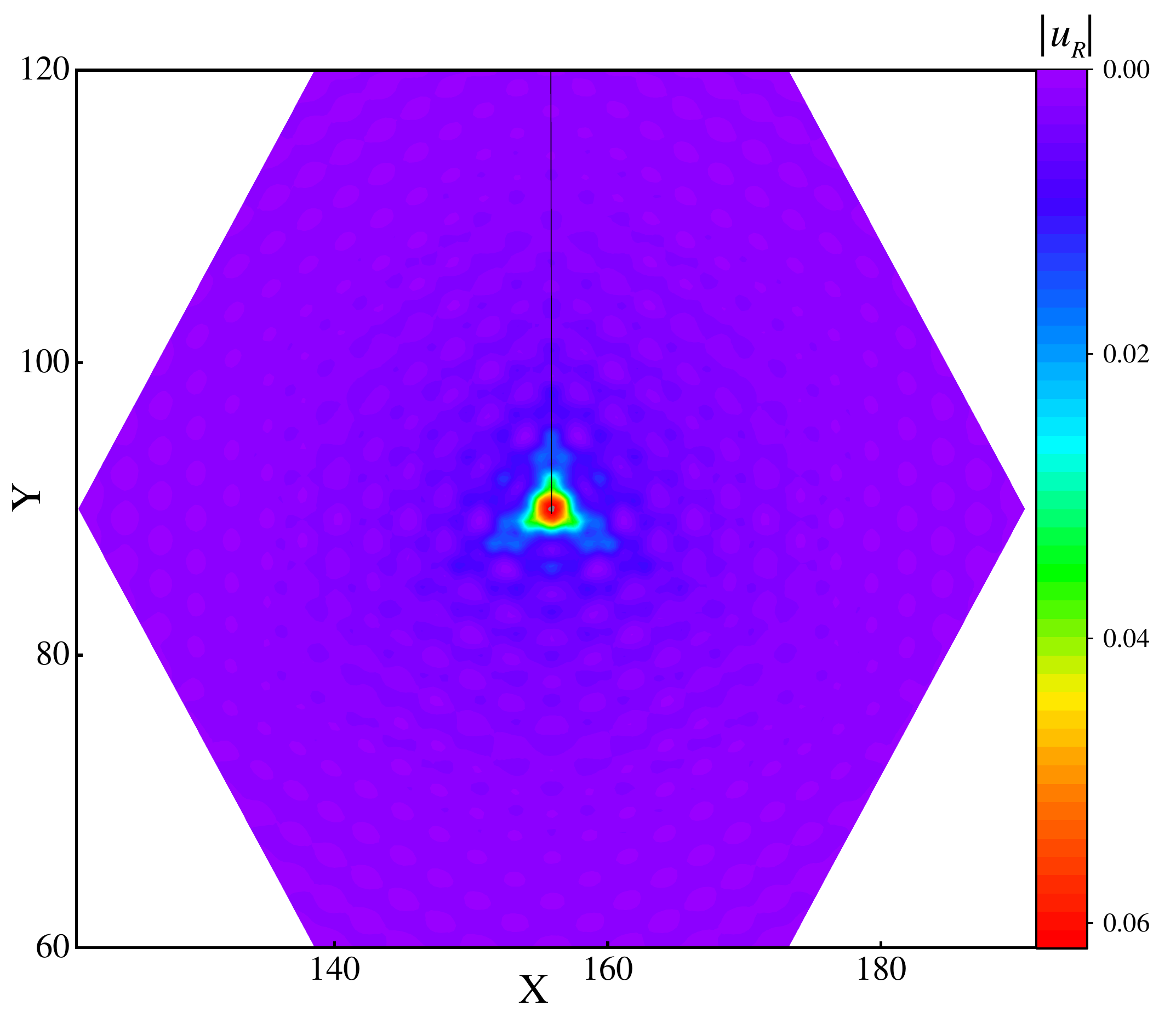}
	\includegraphics[width=7.5cm]{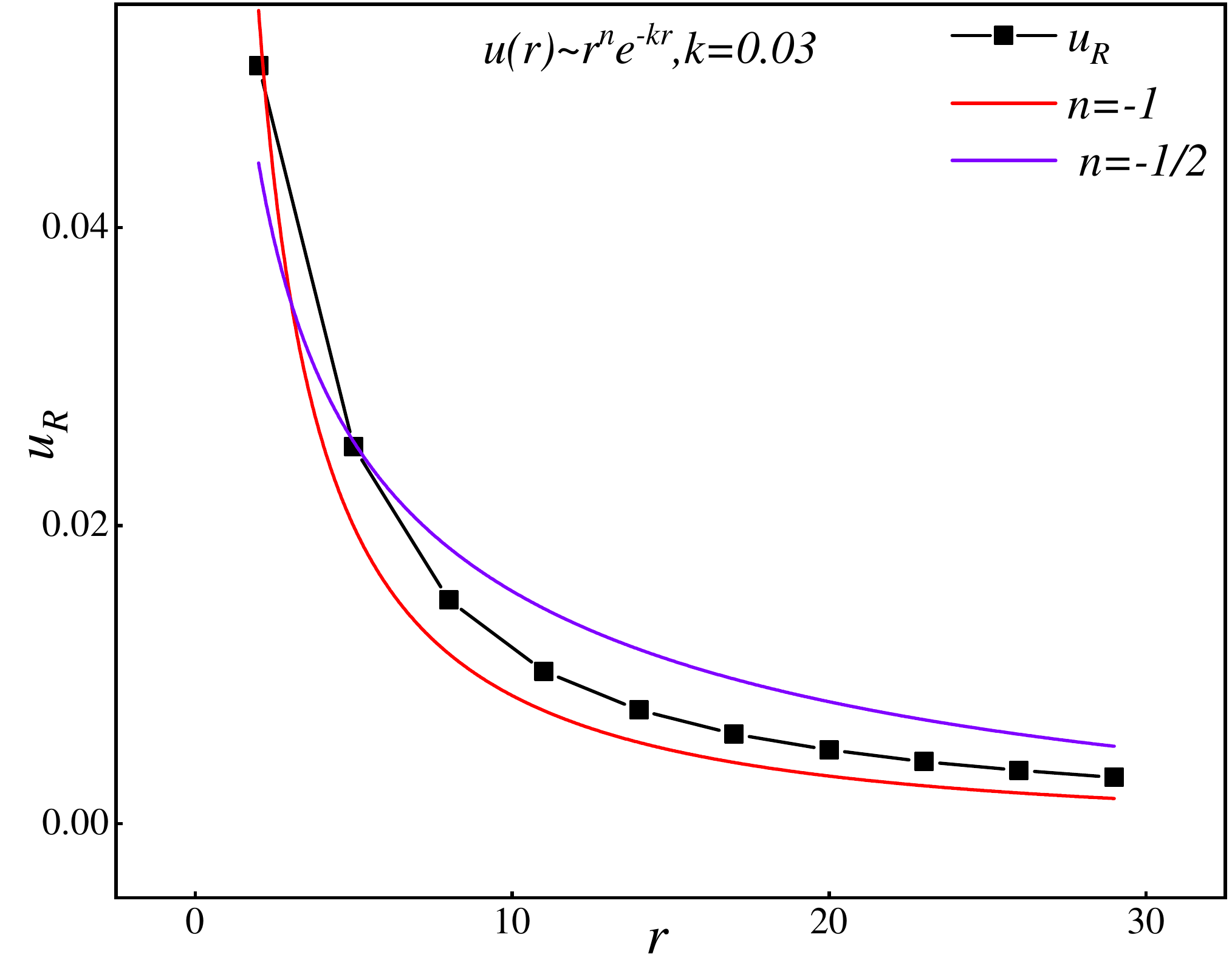}
		\caption{(a)The real part distribution of the amplitude of the zero mode for a $120\times120\times2$  lattice with a flux free vacancy on a B site in the center and a weak uniform $[111]$ magnetic field.  The parameters are $J_x=J_y=J_z=1, t=0.005$. The imaginary part of the amplitude  is the same as the real part.  Only the distribution of the central area including the vacancy is shown. (b) The real part of the amplitude on the A sublattice along the line in (a) as a function of the distance $r$ to the vacancy center. (The unit of $r$ is the bond length of the hexagon.) The amplitude fits the curve $u_R\sim \frac{1}{r} e^{-kr}$, where $k=\Delta/\tilde{J}, \Delta=6\sqrt{3}t, \tilde{J}=\sqrt{3}J$. }\label{fig:zero_mode}
\end{figure}

To solve the above problem, we consider the gapless Kitaev model with finite $t$. In this case, the finite $t$ opens up a gap in the spinon bands with Chern number $1$ or $-1$ and the vacancy induced zero mode decays exponentially in space. Different from the $t=0$ case, the A and B sublattices are now coupled together in the zero mode as shown in Eq.(\ref{eq:A_and_B}) and Eq.(\ref{eq:B_and_A}) so the amplitudes on both sublattices are non-zero. However, the zero mode at finite $t$ is no longer solvable analytically from Eq. (\ref{eq:A_and_B})-(\ref{eq:B_and_A}) and the boundary condition. We then solve the zero mode at finite $t$ by numerically diagonalizing the corresponding bilinear Hamiltonian Eq.(\ref{eq:bilinear_hamiltonian}). The Majorana zero mode $\psi$ we obtained has ${\rm Re \psi= Im \psi}$, i.e., $\psi^\dag=e^{i\pi/2}\psi$, in the flux free case. The phase factor $e^{i\pi/2}$ may be gauged away by transformation $\psi \to \psi e^{-i\pi/4}$, so we only need to study the real part of $\psi$.
The distribution of the real part of the zero mode amplitude in the flux free case is shown in Fig.\ref{fig:zero_mode}a for the isotropic Kitaev model and a $120\times120\times 2$ lattice with a vacancy on the B sublattice in the center of the lattice. The parameters in the plot are $J_x=J_y=J_z=1, t=0.005$. 
We only show the distribution of the zero mode around the central area of the vacancy since the influence from the edge mode  in this area is negligible. Whereas near the edge of the lattice, the amplitude of the zero mode is dominated by the edge mode, which is not our focus in this subsection.

Though it's hard to solve the  exact amplitude distribution of the vacancy induced  zero mode for finite $t$,  at small $t$, which is the case we focus on in this work, we can obtain the asymptotic behavior of the amplitude of the zero mode by analysis. Since the finite  $t$ opens up a gap in the gapless Kitaev model, the amplitude of the zero mode on each sublattice no longer decay algebraically with distance to the vacancy, instead, it decays exponentially with distance to the vacancy center as $a_{lj} \sim e^{-|\Delta r/ \tilde{J}|}$, where $\Delta=6\sqrt{3}t$ is the gap, $\tilde{J}=\sqrt{3}J$ and $r$ is the distance to the vacancy center.  However, the amplitude of the zero mode on the two sublattice depends differently on $t$.  At $t=0$, $a_{ij}$ reduces to Eq.(\ref{eq:amplitude}) so at leading order of $t$, the amplitude on the A sublattice has the form 
\begin{equation}\label{eq:amplitude_finite_t}
a(x, y)\sim {\rm Re} \{ \frac{e^{i q^* x /\sqrt{3}+2i(\pi+\theta^*)y/3}}{(\alpha -2i\beta)y/\sqrt{3}-i x } \}e^{-|\Delta r/ \tilde{J}|}.
\end{equation}
This asymptotic behavior is verified by our numerical result in Fig. \ref{fig:zero_mode}b. In Fig. \ref{fig:zero_mode}b,  we plot the amplitude on the A sublattice Vs. the distance to the vacancy center along the  line in Fig. \ref{fig:zero_mode}a. One can see that the amplitudes on the A sublattice fit very well with the expression $\sim\frac{1}{r}e^{-\Delta r/ \tilde{J}}$. (The small deviation may come from the small amount of mixing from the edge zero mode.) On the B sublattice, the amplitude of the zero mode vanishes at $t=0$. At small finite $t$, the coupling between the A and B sublattice in Eq. (\ref{eq:A_and_B}) and (\ref{eq:B_and_A}) results in an amplitude on the B sublattice proportional to $\sim t/J$ at leading order, as can be seen from Eq. (\ref{eq:B_and_A}).
 The amplitudes on the B sublattice are then much smaller than the amplitudes on the A sublattice in the same unit cell, as can also be seen in the amplitude distribution in Fig.\ref{fig:zero_mode}a. The linear dependence of $b_{lj}$ on $t/J$ is also verified in our numerics by a sign change of $t$, which results in a sign change of $b_{lj}$ but not $a_{lj}$.

\begin{figure}
	\includegraphics[width=7.5cm]{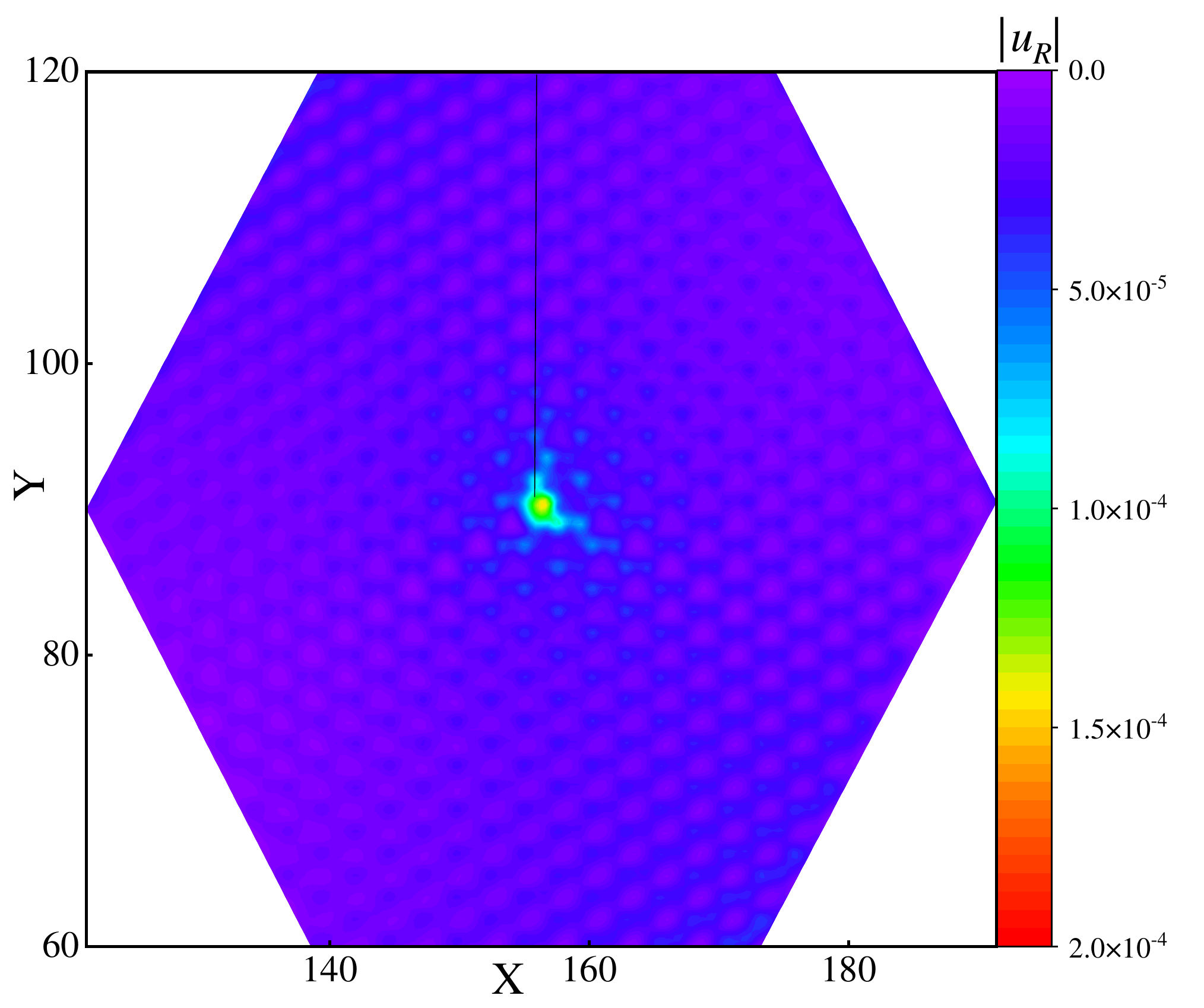}
	\includegraphics[width=7.5cm]{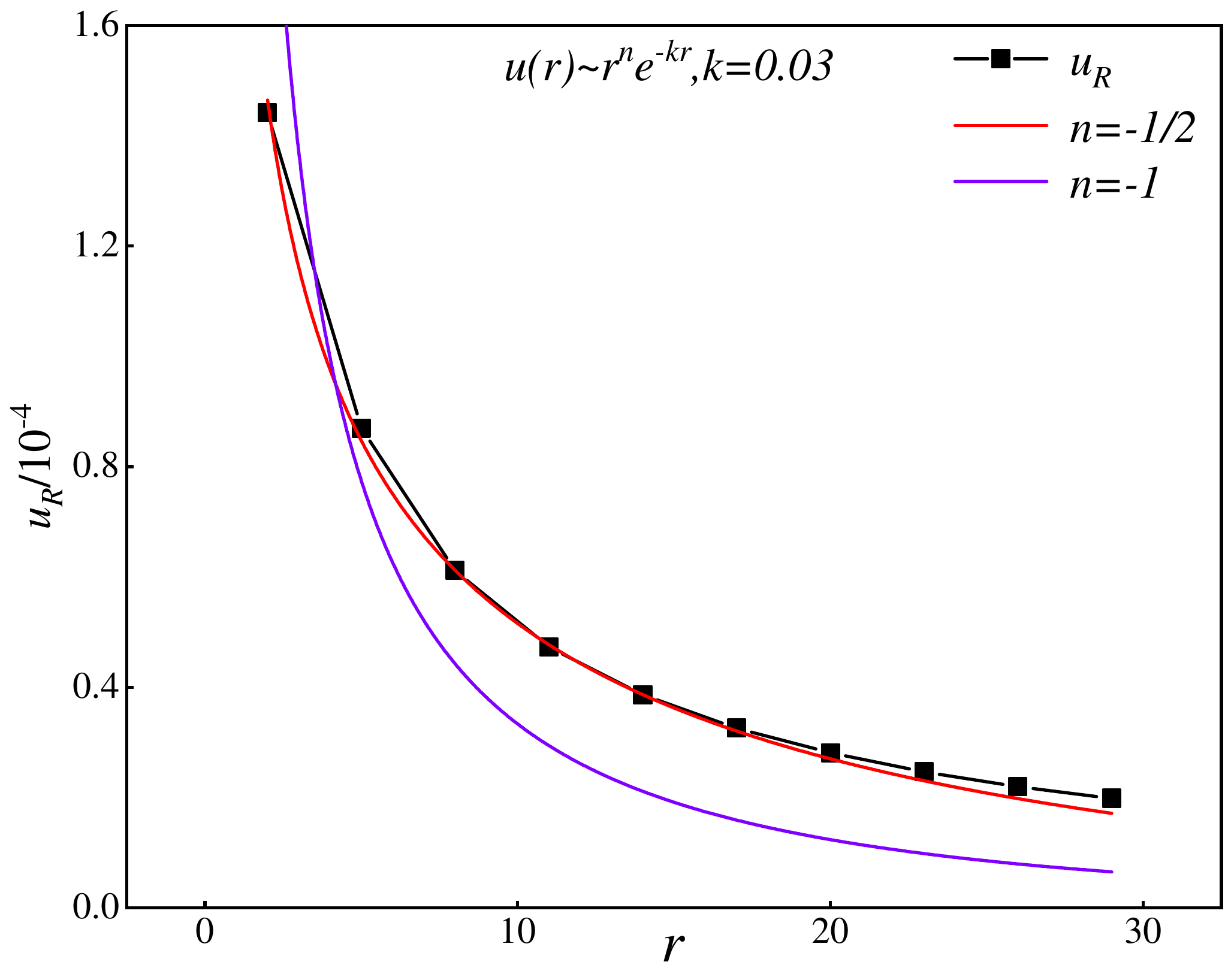}
		\caption{(a)The distribution of the amplitude of the zero mode for a $120\times 120\times2$ Kitaev lattice with a flux threading the vacancy on a B site at the center of the lattice and a weak uniform $[111]$ magnetic field. The parameters are $J_x=J_y=J_z=1, t=0.005$. The amplitude of the zero mode is purely real.  (b)The amplitude of the zero mode on the A sublattice along the line in (a) as a function of the distance to the vacancy center. The amplitude fits the curve $u_R\sim \frac{1}{r^{1/2}} e^{-kr}$, where $k=\Delta/\tilde{J}, \Delta=6\sqrt{3}t, \tilde{J}=\sqrt{3}J$.  }\label{fig:zero_mode_with_flux}
\end{figure}

For the case with a flux threading the vacancy plaquette, the amplitudes of the zero mode also satisfies Eq.(\ref{eq:A_and_B}) and (\ref{eq:B_and_A}) at $l\neq -1, 0, 1$. However, the boundary condition the amplitudes match at the string connecting the vacancy site is different now. Since the z bond operator has the value $-1$ along the half-infinite string at $l=0$, the Eq.(\ref{eq:A_and_B}) and Eq.(\ref{eq:B_and_A}) becomes
\begin{eqnarray}
&&J_x a_{l-1, j+1}+J_y a_{l-1, j} -J_z a_{l, j} -t(b_{l-1,j}-b_{l-1, j+1} \nonumber\\
&&\ \ \ +b_{l, j+1}-b_{l, j-1}-b_{l+1, j-1}+b_{l+1, j})=0, \label{eq:A_and_B_2} \\ 
&&J_x b_{l+1, j-1}+J_y b_{l+1, j} -J_z b_{l, j} -t(-a_{l-1,j}+a_{l-1, j+1} \nonumber\\
&&\ \  \ +a_{l, j+1}-a_{l, j-1}+a_{l+1, j-1}-a_{l+1, j})=0, \label{eq:B_and_A_2}
\end{eqnarray}
for $l=0, j \geq1$. Besides, the boundary condition that Eq.(\ref{eq:A_and_B}) does not exist for $(l, j)=(0, 0)$ and $b_{0, 0}=0$ still needs to be satisfied.

Though the above equations for the zero mode with flux threading the vacancy plaquette is not solvable analytically even at $t=0$, 
the boundary condition near the vacancy still results in nonzero amplitudes only  on the A sublattice (vacancy on the B sublattice) at $t=0$ 
as shown in the appendix.  
At finite $t$ we solve zero mode bound to the flux numerically and compare the asymptotic behavior of the amplitude with the analytical results obtained from the continuum model in the last section.

   In Fig. \ref{fig:zero_mode_with_flux}a, 
 we show the numerical result of the amplitude distribution of the zero mode in the Kitaev model with finite $t$ 
 and a flux threading the vacancy plaquette for a $120\times120\times 2$ lattice. The parameters are $J_x=J_y=J_z=1, t=0.005$. 
 The zero mode we obtained in this case has purely real amplitude, indicating a Majorana zero mode. 
 The plot in Fig. \ref{fig:zero_mode_with_flux}b shows the amplitude on the A sublattice  along the  line in Fig. \ref{fig:zero_mode_with_flux}a. 
 One can see that the amplitudes fit the expression $a(x, y)\sim \frac{1}{r^{1/2}} e^{-\Delta r/ \tilde{J}}$ very well,
 which is consistent with the asymptotic behaviour of the zero mode bound to the $\pi$ flux in the Kitaev model obtained from the continuum model in the last section.
  The amplitude on the B sublattice is also proportional to $t/J$ as can be seen from Eq.(\ref{eq:B_and_A}) and 
  is much smaller than the amplitude on the A sublattice as shown in Fig. \ref{fig:zero_mode_with_flux}a.

From above, we see that  the Majorana zero modes in the flux free and flux threading case  have different asymptotic behaviors  in the Kitaev model with finite $t$. 
As a comparison, we note that for the pure gapped  Kitaev model with a single site vacancy, Ref.~\cite{Willans2011} shows that the amplitudes of the zero mode in both the flux  free and flux threading case are the same.
The reason is because in this case the amplitude of the zero mode vanishes on one of the sublattices. The flux which flips the sign of the bond operator $u_{\langle ij\rangle_\alpha}$ along the half-infinite string then does not affect the zero mode.
On the other hand, the pure gapped Kitaev model has Chern number zero so a vortex does not bind an intrinsic Majorana zero mode. The zero mode in either the flux  free or flux threading case with a single site vacancy is not robust and will disappear if the vacancy includes even number of sites.

\subsection{3.3. Ground state with vortex binding to a vacancy in the lattice Kitaev model}

The flux-threading and flux free case in the last subsection belongs to two independent eigensectors of the Hamiltonian Eq.(\ref{eq:gap_Hamiltonian}). In this subsection, we show that for the Kitaev model in  the weak $[111]$ magnetic field, the ground state with a vacancy  binds a flux in a certain regime of the magnetic field and thus results in a robust Majorana zero mode bound to the vacancy as shown in the last subsection.

\begin{figure}
	\includegraphics[width=7.5cm]{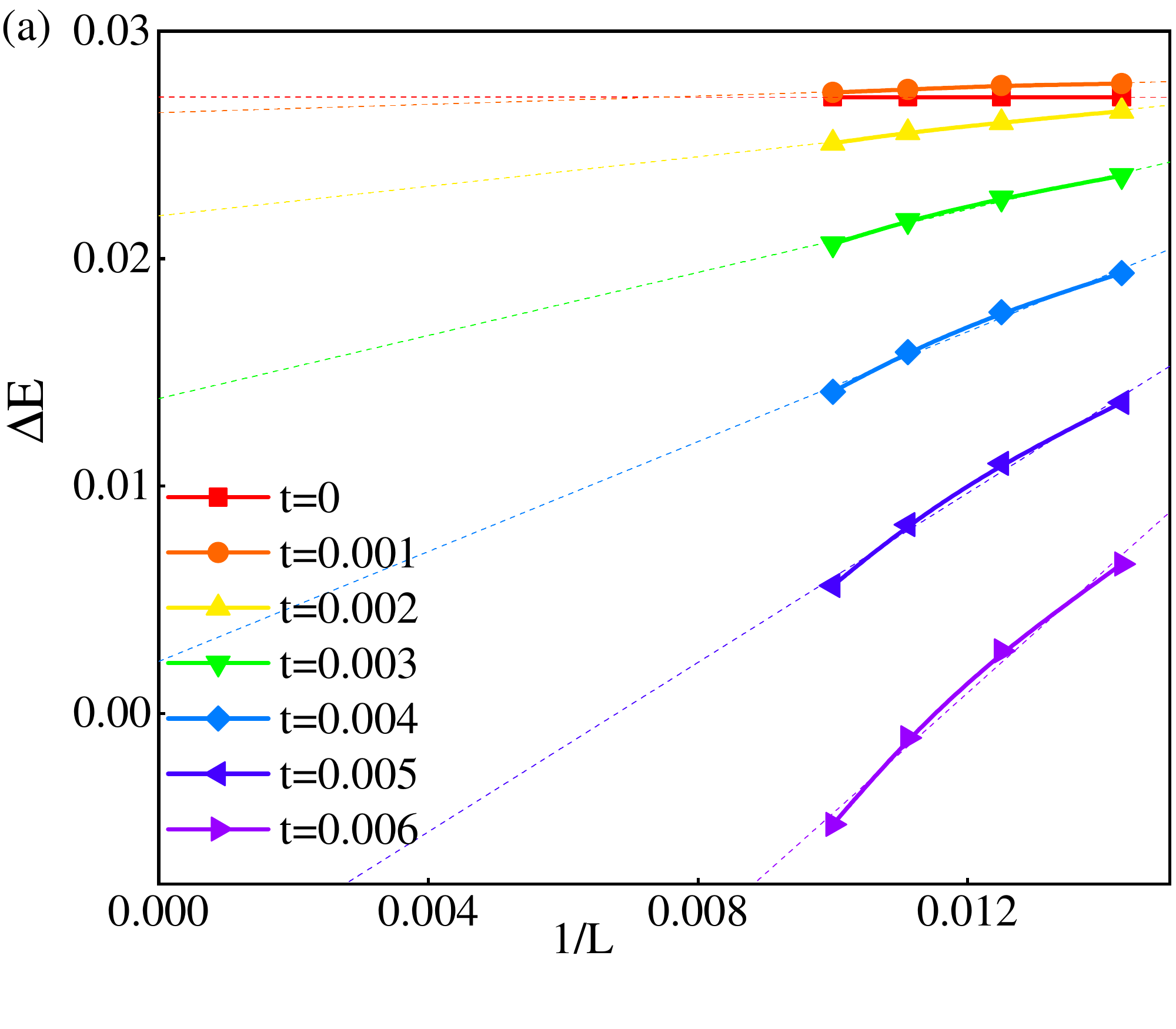}
	\includegraphics[width=7.5cm]{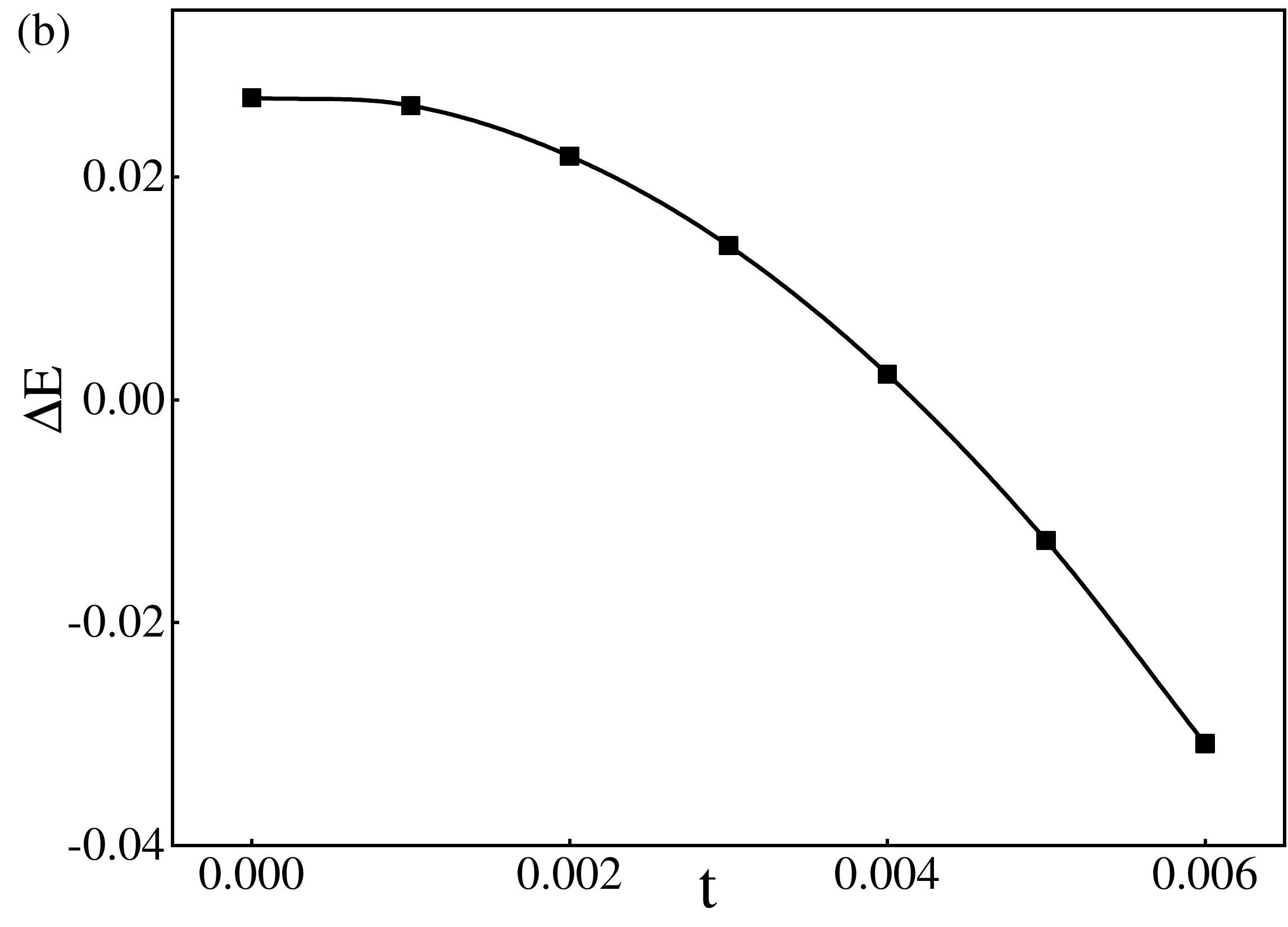}
		\caption{(a)Finite size scaling of the energy difference at $W_I=1$ and $W_I=-1$ of the Hamiltonian Eq.(\ref{eq:GS_Hamiltonian}) with a vacancy. The system size is $L*L*2$, $\Delta E=E_0-E_{{\rm flux}}$, where $E_0$ is the ground state energy with $W_I=1$ and $E_{{\rm flux}}$ is the ground state energy  with $W_I=-1$. (b) The energy difference $\Delta E$ extrapolating to the thermodynamic limit $L\to \infty$ as a function of $t$.  }\label{fig:finite_size_scaling}
\end{figure}

It has been shown in Ref.~\cite{Willans2010, Willans2011} that the ground state of the isotropic gapless Kitaev model with a single vacancy binds a $\pi$ flux and the ground state energy difference between the flux and flux-free case is $-0.027J$. In Fig.\ref{fig:finite_size_scaling}, we show this energy difference  at finite $t$, i.e., when a  weak uniform $[111]$ magnetic field h is applied.
To remain in the perturbative regime so the bilinear Hamiltonian Eq.(\ref{eq:gap_Hamiltonian}) is valid, we assume $t\sim h^3$ is small and $h\ll 1$. (From the numerics in our previous work, $t \approx 2.7h^3$~\cite{Jiang2020}.)
 To minimize the finite size effect, we did finite size scaling of the energy difference between the flux and flux-free case at each point of $t$ in Fig.\ref{fig:finite_size_scaling}a. The energy difference at each $t$  in Fig.\ref{fig:finite_size_scaling}b is the value that extrapolates to the thermodynamic limit $L\to \infty$. From the plot, we can see that the energy gain of the system with the flux threading the vacancy decreases with the increase of the magnetic field $h$ and becomes negative at around $t/J> 0.004$. We then see that for isotropic Kitaev model in a uniform $[111]$ magnetic field, at about $h/J < 0.11$, a single vacancy  binds a flux in the ground state and a Majorana zero mode associated with it. This vortex bound Majorana zero mode decays exponentially with distance to the vacancy and is robust against local perturbations~\cite{Volovik1999} and other Majorana zero modes farther than the decay length  $\xi\sim J/6t$~\cite{Ivanov2001}. These Majorana zero modes may then have potential use in braiding in quantum computing.

\section{V. Majorana zero modes induced by local polarization}

In this section, we study a second type of defect in the Hamiltonian Eq.(\ref{eq:gap_Hamiltonian}), i.e., a locally polarized spin in the uniform system, and the possible Majorana zero mode bound to such defect. The polarized spin is another type of topologically trivial spot and may be achieved and manipulated by a local magnetic field.

\subsection{5.1. Model and Method}

For simplicity, we consider a local conical magnetic field applied to a single site on top of the system described by Hamiltonian Eq.(\ref{eq:gap_Hamiltonian}), i.e., the Kitaev model with a perturbative uniform $[111]$ magnetic field. The total Hamiltonian in the Majorana representation has the following form
\begin{equation} \label{eq:perturbation_Hamiltonian_2}
H=H_K+H_h+H_{loc}, 
\end{equation}
where $H_K$ is the Kitaev Hamiltonian  in Eq.(\ref{eq:ground_state}) and  $H_h$  the third order perturbative term of the weak $[111]$ magnetic field in Eq.(\ref{eq:perturbation}), and
\begin{equation}
H_{loc} = i\sum_\alpha h_{loc}  c_0 b^\alpha_0,
\end{equation}
is a local $[111]$ magnetic field on a site labeled $0$ as shown in Fig.\ref{fig:lattice}c. As a comparison to the single vacancy defect, we assume the site $0$ is also on the B sublattice.

The local constraint $D_i=c_i b^x_i b^y_i b^z_i=1$ is  equivalent to $Q^\alpha_i\equiv c_i b^\alpha_i+\epsilon_{\alpha \beta \gamma} b^\beta_i b^\gamma_i=0$~\cite{Jiang2020}. We impose this constraint by a Lagrangian multiplier in the Hamiltonian:
\begin{equation}
H_\lambda=i\sum_{i, \alpha} \lambda_\alpha (c_i  b^\alpha_i + \epsilon_{\alpha \beta \gamma} b^\beta_i b^\gamma_i).
\end{equation}

In the following, we study the evolution of the the above system with the increase of the local magnetic field and explore possible flux binding in the ground state upon local polarization and the Majorana zero mode associated with the flux.

With only $H_K$ and $H_h$, the bond operator is still conserved, so does the flux in each hexagon plaquette. However, when a local magnetic field is applied on site $0$ shown in Fig.\ref{fig:lattice}c, the flux through the three hexagon plaquettes including the site $0$ is not conserved though the total flux of this three plaquettes is conserved. We call the big plaquette including these three honeycomb plaquettes (the shaded area in Fig.1a) the impurity plaquette and call its flux operator  $W_I$ the impurity flux.

We work in the flux sector that all the outer hexagon have $W=1$. This corresponds to the ground state of the pure Kitaev model as well as the ground state of the Kitaev model with a vacancy at site $0$. Since the impurity flux $W_I$ is conserved, we deal with the two sectors with $W_I=+1$ and $W_I=-1$ separately. Note that for the pure Kitaev model, the ground state has $W_I=+1$, whereas for the Kitaev model with a vacancy at site $0$, the ground state has $W_I=-1$. 

When the local magnetic field is applied on site $0$, only the three bond operators connected to site $0$ are non-conserved. For the impurity flux free case,  we may set all the other bond operators $u_{\langle ij \rangle_\alpha}=i b^\alpha_i b^\alpha_j$ equal to $1$.  For the case with $W_I=-1$, a string with $u_z=-1$ connecting the impurity plaquette and the boundary of the system is introduced as shown in Fig.\ref{fig:lattice}a. The open boundary condition is applied.
Under this convention, the local constraint $D_i=1$ only need to be imposed on site $i=0$.

For convenience, we relabel the three nearest neighbor sites connected to site $0$ as $\alpha=x, y, z$ as shown in Fig.\ref{fig:lattice}c, and the six next nearest neighbor sites connected to site $0$ as $\alpha_r$, where $x_r=1,2, y_r=3,4, z_r=5,6$ are labeled in Fig. \ref{fig:lattice}c respectively.

The Kitaev Hamiltonian then becomes 
\begin{equation}
H_K=i\sum_{\langle ij\rangle_\alpha'} J_\alpha c_i c_j -\sum_\alpha J_\alpha b^\alpha_0 b^\alpha_\alpha c_0 c_\alpha, 
\end{equation}
where $\langle ij\rangle_\alpha'$ represents all the bonds except the three nearest neighbor bonds connected to site $0$.

The Hamiltonian $H_h$ becomes
\begin{eqnarray}
H_h &=& i t \sum_{\langle\langle ik\rangle\rangle'} c_i c_k \nonumber\\
&&-t \sum_{\alpha, \gamma} i c_\alpha b^\alpha_\alpha (i\epsilon_{\alpha\beta\gamma} b^\alpha_0 b^\gamma_0) ic_\gamma b^\gamma_\gamma \nonumber\\
&&-t \sum_{\alpha, \gamma} i c_0 b^\alpha_0 (i\epsilon_{\alpha\beta\gamma} b^\alpha_\alpha b^\gamma_\alpha) ic_{\alpha_r} b^\gamma_{\alpha_r}, 
\end{eqnarray}
where $\langle\langle ik\rangle\rangle'$ represents the next nearest neighbors except those involving the impurity site $0$, i.e., none of the three sites $i, j, k$ in Fig.\ref{fig:lattice}b is on site $0$.
 The second term in $H_h$  represents the term $\sim \sigma^\alpha_\alpha \sigma^\beta_0 \sigma^\gamma_\gamma$ and the third term represents $\sim \sigma^\alpha_0 \sigma^\beta_\alpha \sigma^\gamma_{\alpha_r}$. Here the sites $\alpha, \gamma=x, y,z$ in the sum correspond to two of the three sites connected to the site $0$ as shown in Fig. \ref{fig:lattice}b and $\alpha\neq \gamma$.

 We then use the mean field theory to study the local polarization process  in the following. The applicability of the mean field theory has been tested in previous work on a $[001]$ local magnetic field acting on the Kitaev model, where the mean field results agree very well with the exact numerical renormalization group results~\cite{Liang2018}. Actually the mean field theory is exact at $h_{loc}\to \infty$ and is especially relevant at the locally polarized phase where the fluctuation of bond operators connected to site $0$ is highly suppressed. And this is the regime we are especially interested in in this work.

We   decouple the quartic terms in $H_K$ and $H_h$ to quadratic terms by the mean field theory as follows:
\begin{eqnarray}
H^{MF}_K &=& i\sum_{\langle ij\rangle_\alpha'} J_\alpha c_i c_j + \nonumber\\
&& \sum_\alpha [J_\alpha \langle i b^\alpha_0 b^\alpha_\alpha\rangle ic_0 c_\alpha +J_\alpha \langle i c_0 c_\alpha \rangle i b^\alpha_0 b^\alpha_\alpha \nonumber\\
&& -J_\alpha \langle i c_0 b^\alpha_0 \rangle i c_\alpha b^\alpha_\alpha -J_\alpha \langle i c_\alpha b^\alpha_\alpha\rangle i c_0 b^\alpha_0 ],
\end{eqnarray}
and 
\begin{eqnarray}
H^{MF}_h&=&i t \sum_{\langle\langle ik\rangle\rangle'} c_i c_k -t \sum_{\alpha\gamma} \epsilon_{\alpha\beta\gamma} \nonumber\\
&&\{ [\langle i b^\alpha_0 b^\alpha_\alpha\rangle \langle i b_0^\gamma b^\gamma_\gamma \rangle ic_\alpha c_\gamma+\langle i b^\alpha_0 b^\alpha_\alpha\rangle \langle  ic_\alpha c_\gamma \rangle  i b_0^\gamma b^\gamma_\gamma \nonumber\\
&&+\langle  i b_0^\gamma b^\gamma_\gamma \rangle \langle  ic_\alpha c_\gamma \rangle  i b^\alpha_0 b^\alpha_\alpha ] \nonumber\\
&&+[ \langle i c_\alpha b^\alpha_\alpha \rangle \langle ic_0 b^\beta_0\rangle i c_\gamma b^\gamma_\gamma
+\ \langle i c_\alpha b^\alpha_\alpha \rangle \langle i c_\gamma b^\gamma_\gamma\rangle  ic_0 b^\beta_0 \nonumber\\
&&+ \langle  ic_0 b^\beta_0\rangle \langle i c_\gamma b^\gamma_\gamma\rangle   i c_\alpha b^\alpha_\alpha] \nonumber\\
&&+ [ \langle i b^\alpha_0 b^\alpha_\alpha \rangle \langle i b^\gamma_{\alpha_r} b^\gamma_\alpha \rangle ic_0 c_{\alpha_r}
+\langle i b^\alpha_0 b^\alpha_\alpha \rangle \langle ic_0 c_{\alpha_r}  \rangle i b^\gamma_{\alpha_r} b^\gamma_\alpha \nonumber\\
&&+\langle i b^\gamma_{\alpha_r} b^\gamma_\alpha \rangle \langle ic_0 c_{\alpha_r}  \rangle   i b^\alpha_0 b^\alpha_\alpha]\}.
\end{eqnarray}
In the above decoupling, we neglect all the terms with $\langle i b^\alpha_i c_j\rangle$,  $i\neq j$ since the mean field value of this average is zero from previous study~\cite{Liang2018, Liang2018_2, Jiang2020}. The bond operator $\langle i b^\gamma_{\alpha_r} b^\gamma_\alpha \rangle$ is still conserved so we  set it to be one. 

There are then altogether $24$ mean field parameters to be determined self-consistently, which are

 $\langle i c_0 b^\alpha_0 \rangle, \langle i c_\alpha b^\alpha_\alpha \rangle$, $\alpha=x, y, z$; 
 
$ \langle i b^\alpha_0 b^\alpha_\alpha \rangle, \langle i c_0 c_\alpha\rangle, \alpha=x, y,z;$

$ \langle i c_\alpha c_\gamma \rangle, \langle i c_0 c_{\alpha_r}\rangle, \alpha, \gamma=x, y,z$ and $\alpha_r=1, 2,..., 6$;

and three Lagrangian multipliers $\lambda_\alpha $ on site $0$.

The above mean field parameters are determined self-consistently by iteration. The Lagrangian multipliers are determined by  iterating the initial value of $\lambda_\alpha$ and the mean field parameters until $\langle Q^\alpha_0\rangle=0$.

\subsection{5.2. Mean field results of the magnetization process}

\begin{figure}
	\includegraphics[width=8cm]{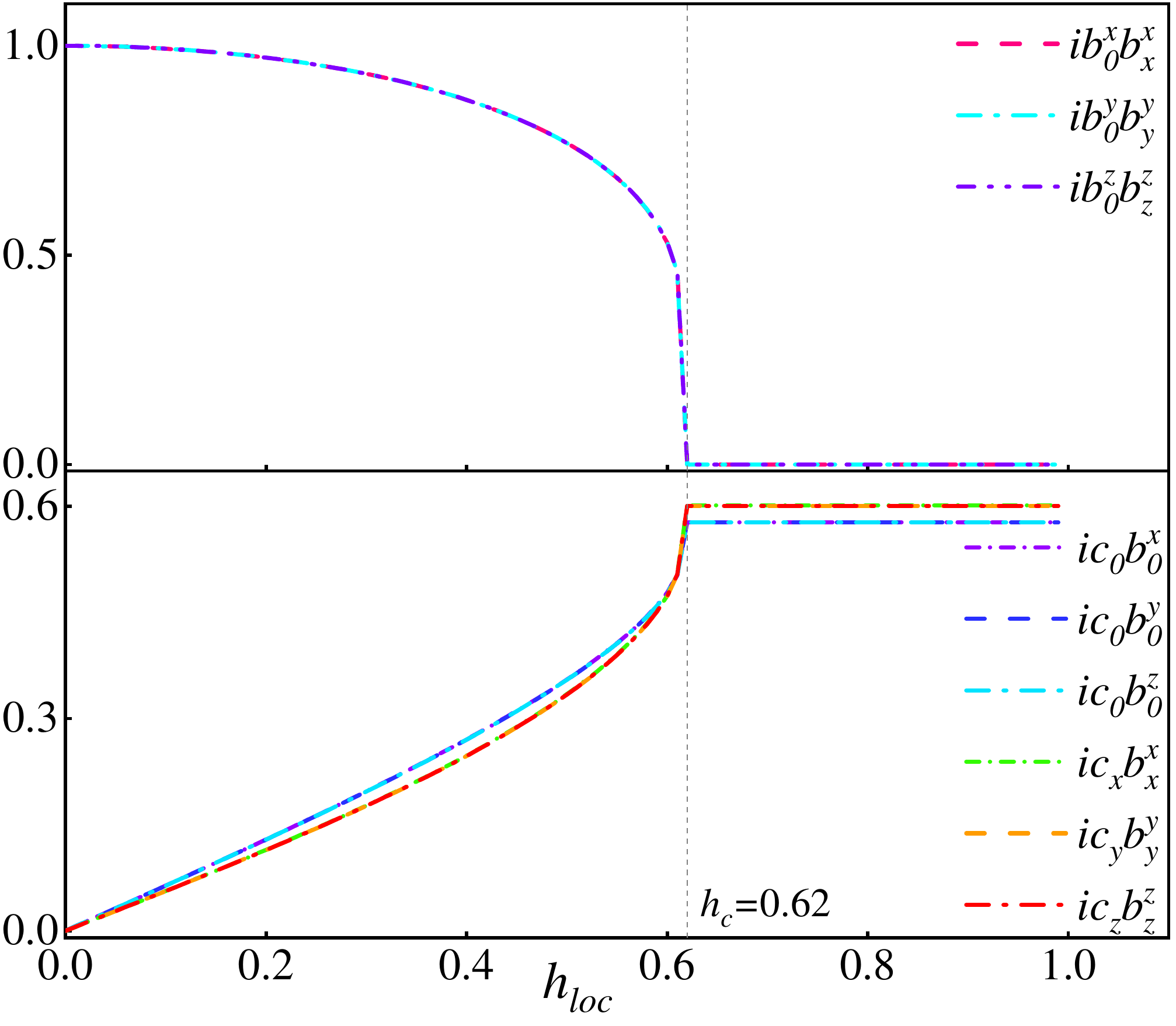}
		\caption{The evolution of the mean field parameters as a function of the local magnetic field on site $0$ in the impurity flux free case for a $64*64*2$ lattice. The parameters $J_x=J_y=J_z=1, t=10^{-4}$. Upper: the three bond operators $u_{\langle 0\alpha\rangle}, \alpha=x, y, z$ connecting site $0$ and site $\alpha$. The mean field value of the three bond operators all goes to zero at the phase transition $h_{loc}=h_c$.  Lower: the local magnetization on site $0$ and site $x, y, z$. After the phase transition at $h_c$, the spin on site $0$ is fully polarized and has $\langle \sigma^\alpha_0\rangle=1/\sqrt{3}$ for all three component $\alpha=x, y, z$.}\label{fig:parameters_noflux}
\end{figure}

\begin{figure}
\includegraphics[width=7.5cm]{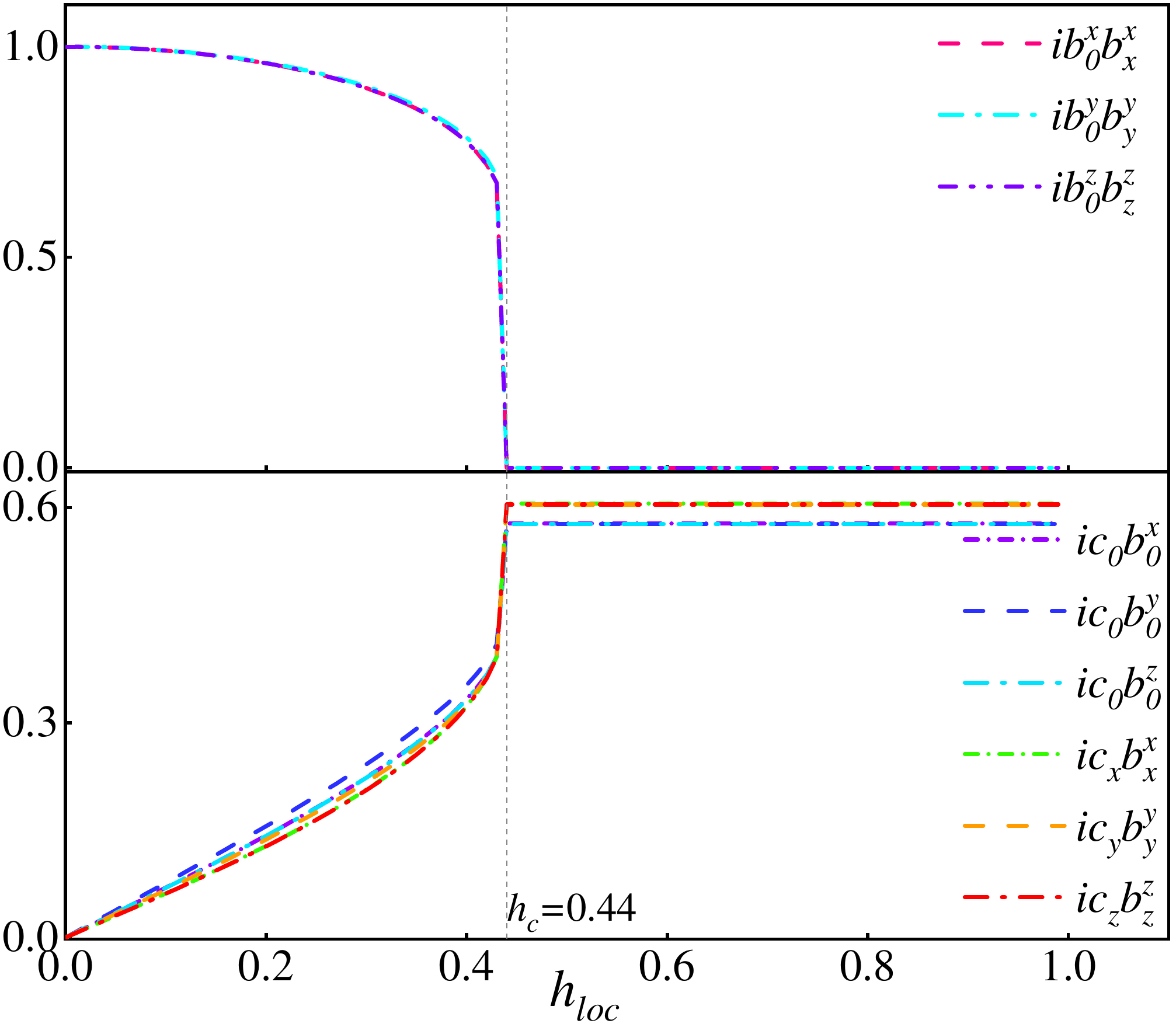}
\caption{
The evolution of the mean field parameters as a function of the local magnetic field on site $0$ in the case with impurity flux $W_I=-1$ for the same lattice in Fig.\ref{fig:parameters_noflux}. The legend is the same as in Fig.\ref{fig:parameters_noflux}. The three bond operators  $u_{\langle 0\alpha\rangle}, \alpha=x, y, z$ also goes to zero after the phase transition at $h_c$ and  the spin on site $0$ is fully polarized with $\langle \sigma^\alpha_0\rangle=1/\sqrt{3}$ for all three components $\alpha=x, y, z$.
}\label{fig:parameters_flux}
\end{figure}

We focus on the isotropic case with $J_x=J_y=J_z=1$. For a given t, the evolution of the mean field parameters for a $64*64*2$ lattice with the increase of the local magnetic field is shown 
in Fig.\ref{fig:parameters_noflux} and Fig.\ref{fig:parameters_flux}  for $W_I=1$ and $W_I=-1$ respectively. The local magnetic field is applied on the B site of the center of the lattice (labeled site $0$). The magnetization on site $0$ shows that at $h_{loc}=h_c$, a first order phase transition~\cite{Note} takes place and the spin on site $0$ is fully polarized with $\langle \sigma^\alpha_0 \rangle=1/\sqrt{3}$ for all three components $\alpha=x, y, z$ after the phase transition. At the phase transition, the mean field value of the three bond operators $u_{\langle 0\alpha\rangle_\alpha}$ connected to site $0$ becomes zero for both $W_I=\pm1$, which results in an effective decoupling of the site $0$ with the neighboring sites.  At the same time, the polarized spin on site $0$ results in three effective local magnetic field $h_0^x, h_0^y, h_0^z$ acting on $\sigma^\alpha_\alpha$ at site $\alpha=x, y, z$ respectively with value  $h_0^\alpha=J_\alpha\langle \sigma_0^\alpha\rangle=J/\sqrt{3}, \alpha=x, y,z$. This results in  finite magnetization $\langle \sigma^\alpha_\alpha\rangle$ on the three $\alpha$ sites as shown in Fig.\ref{fig:parameters_noflux}. All the other magnetization $\langle \sigma^\alpha_i \rangle$ with $i\neq 0, \alpha$ is zero in the lattice.  The polarized phase  may then be described by a bilinear Hamiltonian 
\begin{eqnarray}\label{eq:effective_Hamiltonian}
H &=& i\sum_{\langle ij\rangle'} J_\alpha u_{\langle ij\rangle_\alpha} \hat{c}_i \hat{c}_j+
 t \sum_{\langle\langle ij \rangle\rangle'} i\epsilon_{\alpha\beta\gamma} u_{{\langle ik\rangle}_\alpha} u_{{\langle kj\rangle}_\gamma}\hat{c}_i \hat{c}_j \nonumber\\
&&+\sum_{\alpha=x, y, z} i h^\alpha_0 \hat{b}^\alpha_\alpha \hat{c}_\alpha, \nonumber\\
\end{eqnarray}
i.e., the Hamiltonian Eq.(\ref{eq:gap_Hamiltonian}) with a vacancy on site $0$ and three local effective magnetic field $h^\alpha_0$ acting on  $\hat{\sigma}^\alpha_\alpha$ respectively. This Hamiltonian has been studied at $t=0$ in Ref~\cite{Willans2011} for the response of the vacancy to Zeeman field in the Kitaev model. We will use this Hamiltonian with finite $t$ to study the phase after polarization.

\begin{figure}
\includegraphics[width=8cm]{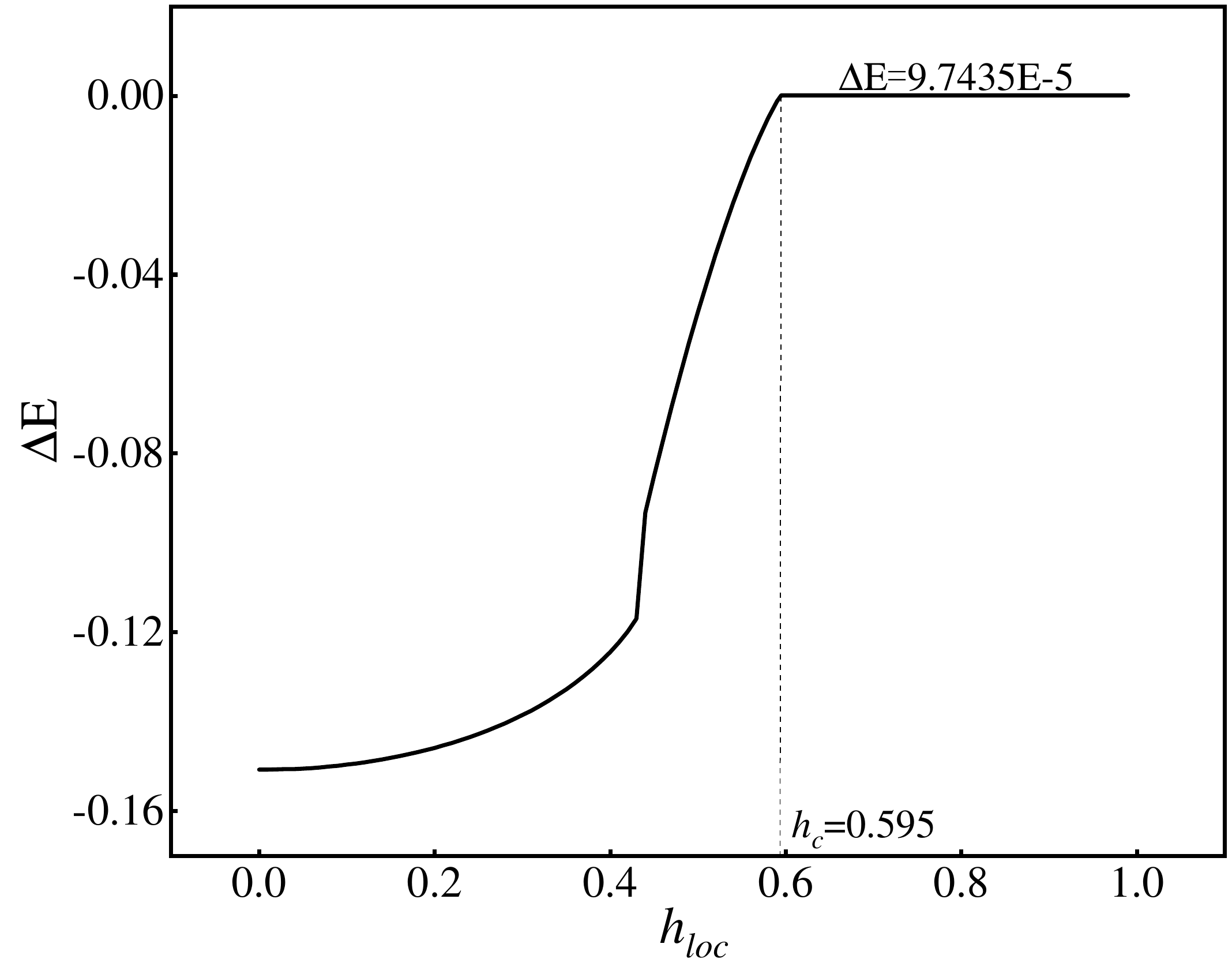}
\caption{The energy difference $\Delta E=E_0-E_{\rm flux}$ between the impurity flux  case and flux free case as a function of the local magnetic field from the mean field theory at $J_x=J_y=J_z=1, t=0.0001$ for a $64*64*2$ lattice. Before the phase transition at $h_{loc}=h_c=0.595$, the flux free case has lower energy. After the phase transitions for both the $W_I=\pm 1$ cases, the  flux state $W_I=-1$ has lower energy, indicating a flux binding at the ground state after the polarization of the spin on site $0$. Note that the transition field $h_c$ as well as $\Delta E$ in this plot is the corrected value after considering superheating of the first order phase transition so $h_c$ is slightly lower than the polarization field $h_c$ in Fig.\ref{fig:parameters_noflux}. See Ref
\cite{Liang2018} for  the correction procedure.}\label{fig:DeltaE_H}
\end{figure}

Figure \ref{fig:DeltaE_H} shows the energy difference of the two cases with $W_I=1$ and $W_I=-1$ as a function of the local magnetic field obtained from the mean field theory for $J=1, t=10^{-4}$ and the lattice with size $64*64*2$. We can see that  the flux free case $W_I=1$ has lower energy in the local magnetic field $h_{loc}$ before the phase transition at $h_{loc}=h_c$. However, after the phase transition, the state with $W_I=-1$ has lower energy though the energy gain of the flux state in this case is much smaller than the case with only a vacancy in the last section. We then see that upon local polarization, the impurity plaquette binds a flux in the ground state at small $t$.

\begin{figure}
\includegraphics[width=7.5cm]{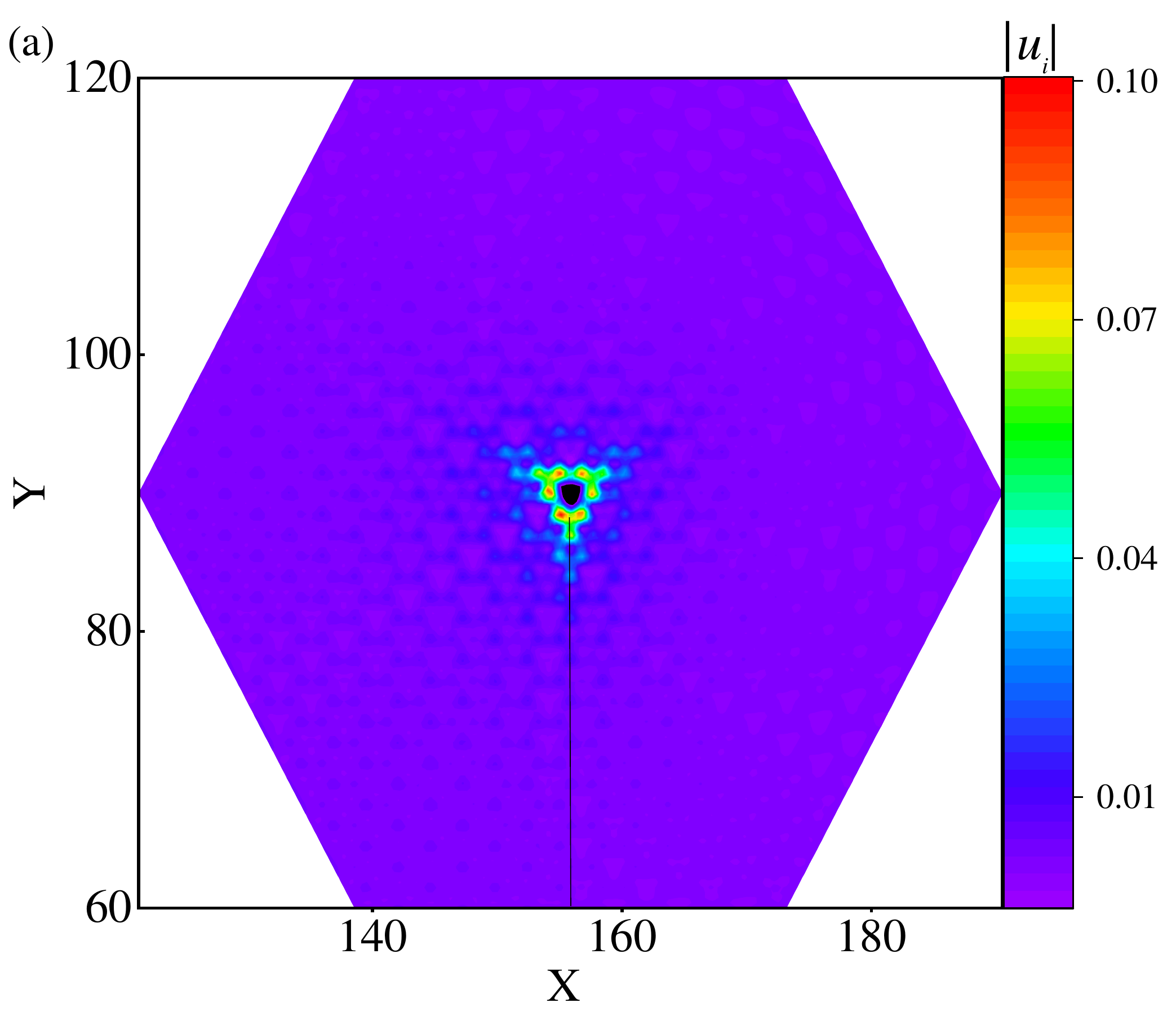}
\includegraphics[width=7.5cm]{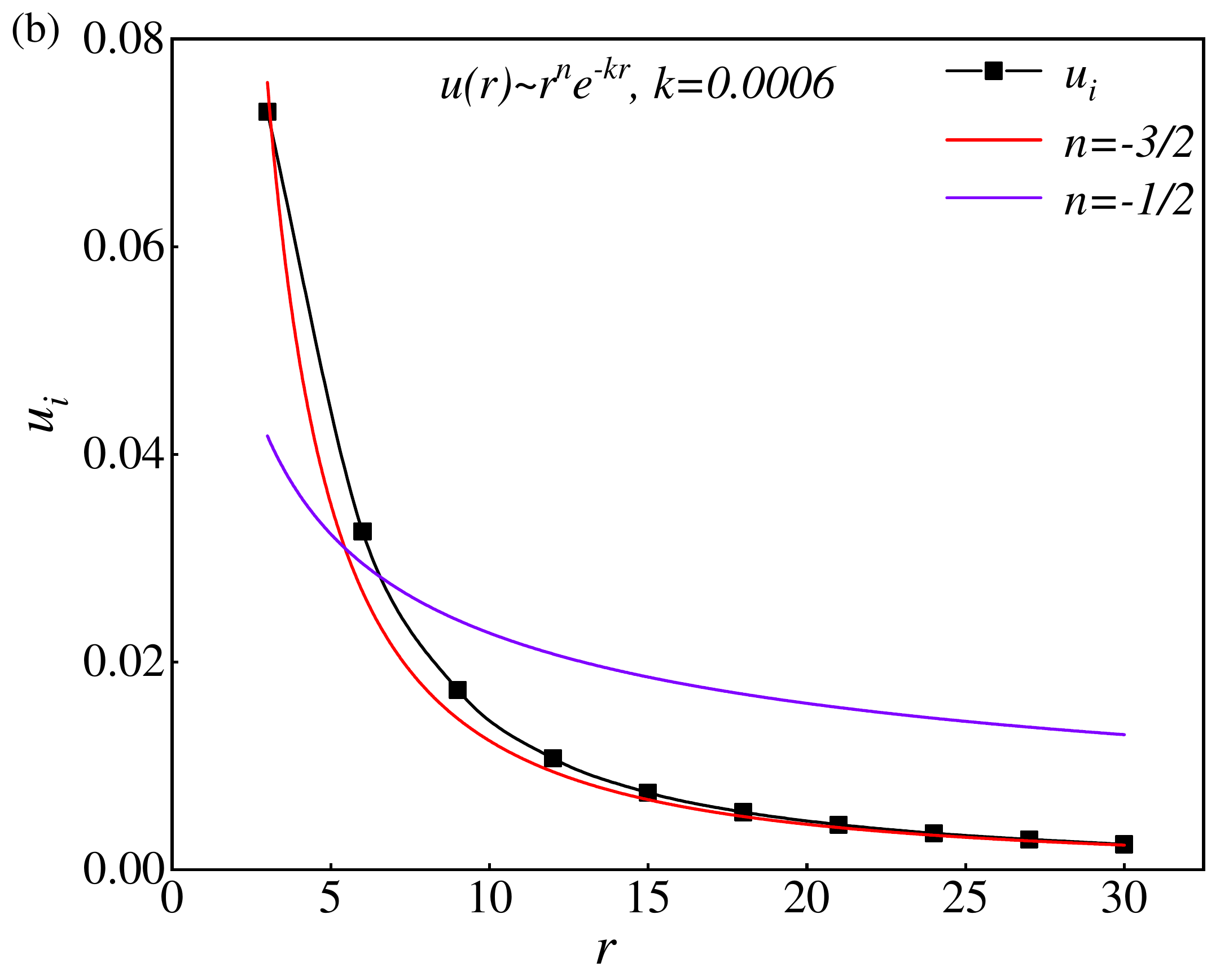}
\caption{(a) The amplitude distribution of the Majorana zero mode bound to the flux after the polarization of a local spin on a $B$ site. The parameters are $J_x=J_y=J_z=1, t=10^{-4}.$ (b)Fitting curve of the amplitude on the B sublattice along the line in  (a). }\label{fig:distribution_local_field}
\end{figure}

The flux threading the impurity plaquette at the ground state after phase transition at $h_c$ results in a Majorana zero mode at the ground state. We obtained a Majorana zero mode with purely imaginary amplitude. The distribution of the imaginary part of the amplitude  is shown in Fig.\ref{fig:distribution_local_field} for $J=1, t=10^{-4}$ and  a local magnetic field on a B  site. Opposite to the vortex induced  zero mode with a vacancy on site $B$ in Fig.\ref{fig:zero_mode_with_flux}, the amplitude of the zero mode on the A sublattice $a_{lj}$ is now much smaller than the amplitude on the B sublattice   as shown in Fig.\ref{fig:distribution_local_field}a.  This is because with the three effective local magnetic field on the three  $\alpha=x, y, z$ site after the polarization, the boundary condition of the zero mode near the site $0$ is different from the case with only a vacancy on site $0$. At $t=0$, the new boundary condition results in a zero mode with  finite amplitude  $b_{ij}$ on the sublattice B but zero amplitude $a_{ij}$ on sublattice A, as shown in the Appendix. This is opposite to the case with only a vacancy on site $B$. (Note that the amplitudes of the three field $\hat{b}^\alpha_\alpha$ on the three A sites $\alpha=x, y, z$ are still nonzero in this zero mode.) For finite small $t$, the coupling between the A and B sublattice in the zero mode equation results in a small amplitude on the A sublattice proportional to $\sim t/J$, as shown in Fig.\ref{fig:distribution_local_field}a. 

The plot in Fig.\ref{fig:distribution_local_field}b shows the amplitude on the B sublattice as a function of the distance to the local polarized site $0$. The amplitude still fits the curve $u(r)\sim \frac{1}{r^a}e^{-\Delta r/ \tilde{J}}$ very well, but with $a=\frac{3}{2}$ instead of $\frac{1}{2}$ for the flux induced Majorana zero mode in Fig.\ref{fig:zero_mode_with_flux}, indicating faster decay in space. An analytic explanation of this difference remains to be explored in future work.
However, the exponential decay of the zero mode guarantees that two such Majorana zero modes far apart have little influence on each other.

For the flux free case, due to the three effective local magnetic field,  three Majorana fermion fields $\hat{b}^\alpha_\alpha$ enter the bilinear Hamiltonian in Eq.(\ref{eq:effective_Hamiltonian}) and the dimension of the bilinear Hamiltonian matrix becomes even  as $(2N+2)*(2N+2)$. There is no Majorana zero mode after polarization  in this case. This is  verified in the spectrum we obtained numerically.

\begin{figure}
\includegraphics[width=7.2cm]{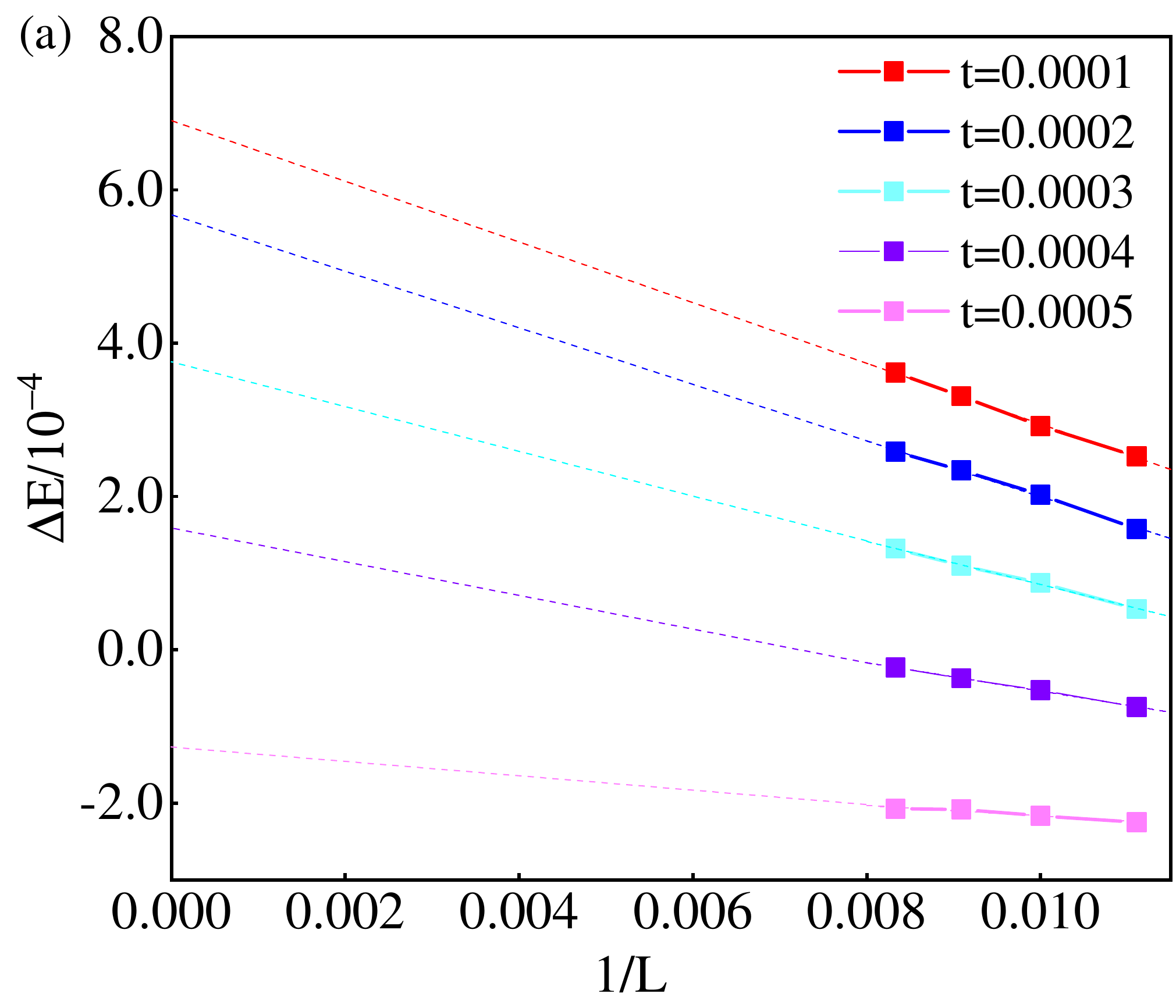}
\includegraphics[width=7.5cm]{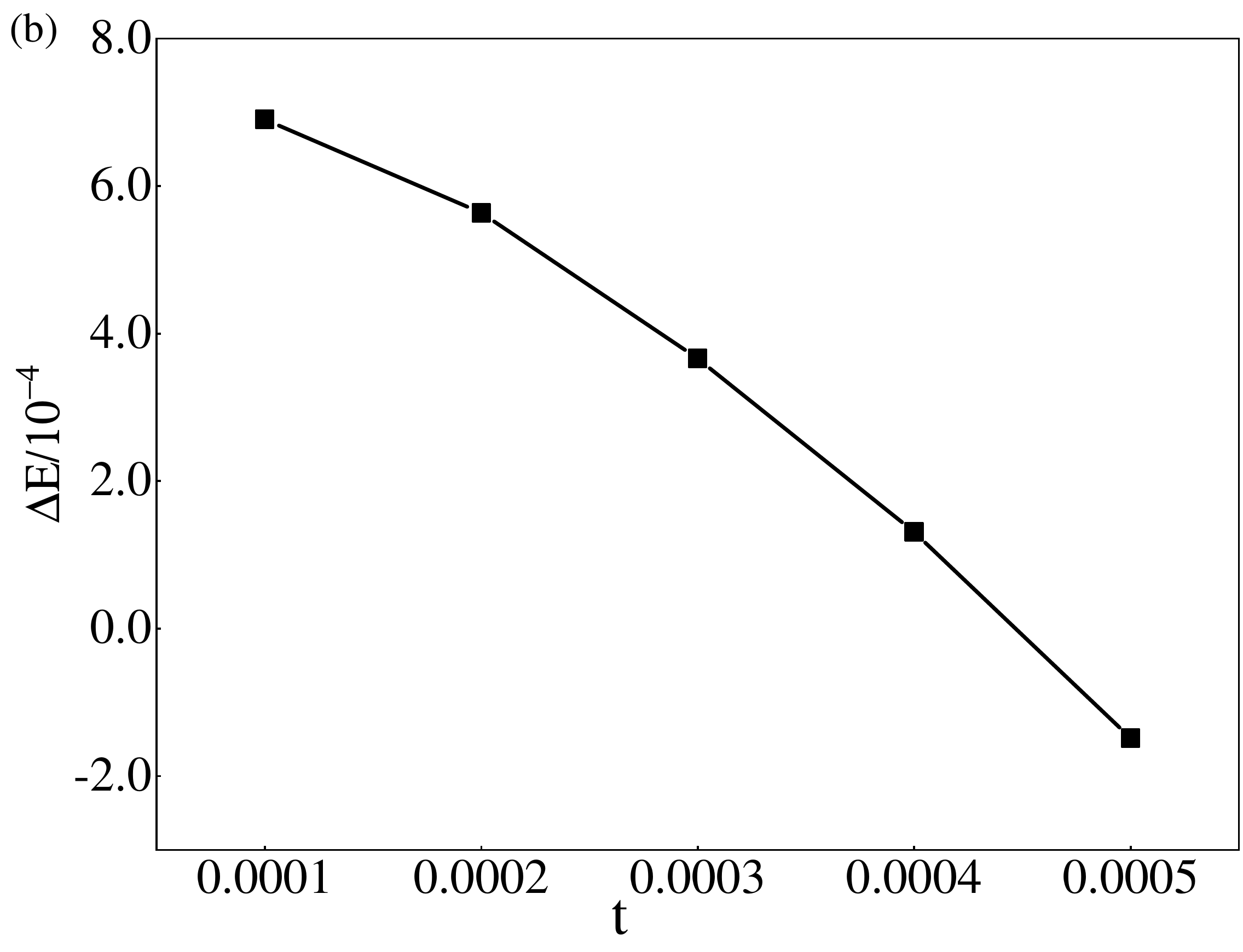}
\caption{(a)Finite size scaling of the  energy difference $\Delta E=E_0-E_{\rm flux}$  between the impurity flux case and the flux free case after the local spin polarization on site $0$ in both cases  for parameters $J_x=J_y=J_z=1, t=10^{-4}$. Both $E_0$ and $E_{\rm flux}$ is computed through the bilinear Hamiltonian Eq.(\ref{eq:effective_Hamiltonian}). (b) The energy difference $\Delta E$ as a function of $t$. $\Delta E$ takes the value extrapolating to $L\to \infty$ in (a).}\label{fig:DeltaE_t}
\end{figure}

With the increase of $t$, however, the energy gain of the flux state after polarization of the local spin at site $0$ becomes negative. As shown in Fig.\ref{fig:DeltaE_t}, at $t>4\times10^{-4}$ and $J=1$, the extrapolation of $\Delta E$ to $L\to  \infty$ becomes negative and the ground state of the polarized state remains  to be the  flux free state $W_I=1$ and there is no Majorana zero mode in the ground state after the polarization.

In summary, we see that by polarizing a local spin in the Kitaev model with a weak uniform $[111]$ magnetic field, the ground state of the system binds a $\pi$-flux  around the polarized spin which results in a Majorana zero mode that decays exponentially in space and is robust against local perturbation. At the same time, it may be easily manipulated by the local magnetic field in space. For these reasons, it may be a potential candidate for braiding in topological quantum computing.

\section{V. Discussions and conclusions}

As a comparison to the Majorana zero modes bound to local impurities in the above sections, we have a brief discussion of the edge Majorana modes for the above Kitaev model in the $[111]$ magnetic field with a straight edge. 

As an analog to the $p$-wave superconductor~\cite{Read2000}, the Kitaev model Hamiltonian Eq.(\ref{eq:GS_Hamiltonian}) has a chiral edge Majorana mode for the system with a straight edge in the continuum limit. One can solve this chiral edge Majorana mode from 
 Eq.(\ref{eq:continuum_Hamiltonian}) and the block diagonalized Hamiltonian ${\cal H}_D$ and ${\cal{H}_{-D}}$. Assume the  wave function of the chiral edge mode as $\Phi_{CM}=u_c\phi_A+iv_c\phi_B+\tilde{u}_c\phi^\dag_A+i\tilde{v}_c\phi^\dag_B$, the coefficients satisfy the equations
\begin{eqnarray}
E u_c&=&\Delta(\textbf{r}) u_c +i \sqrt{3} J (\frac{\partial }{\partial x} + i \frac{\partial }{\partial y} ) v_c , \nonumber\\
 E v_c&=&- \Delta(\textbf{r}
) v_c +i \sqrt{3} J (\frac{\partial }{\partial x} - i \frac{\partial }{\partial y} ) u_c,\nonumber\\ 
E \tilde{u}_c&=&-\Delta(\textbf{r}) \tilde{u}_c +i \sqrt{3} J (-\frac{\partial }{\partial x} + i \frac{\partial }{\partial y} ) \tilde{v}_c , \nonumber\\
 E \tilde{v}_c&=& \Delta(\textbf{r}
) \tilde{v}_c +i \sqrt{3} J (-\frac{\partial }{\partial x} - i \frac{\partial }{\partial y} ) \tilde{u}_c.
\end{eqnarray}
Assume there is an infinite edge along the $y$ direction and the wave function contains the phase factor $e^{ik_y y}$, the above equation set has a special solution at energy $E=-\sqrt{3}J k_y$ and $u_c=i v_c, \tilde{u}_c=i\tilde{v}_c$ for $\Delta/J>0$ with  $u_c \sim\tilde{u}_c \sim e^{-\int \Delta dx/\sqrt{3}J}$,  or $E=\sqrt{3}J k_y$ and $u_c=-iv_c, \tilde{u}_c=-i\tilde{v}_c$ with $u_c\sim \tilde{u}_c \sim e^{\int \Delta dx/\sqrt{3}J}$ for $\Delta/J<0$, where $x$ is the distance to the edge. To get a Majorana mode, we choose $\tilde{u}_c=u^*_c, \tilde{v}_c=-v^*_c$. The above mode then satisfies $\Phi_{CM}=\Phi_{CM}^\dag$ and is thus a Majorana mode. This mode propagates along a single direction of the edge so is chiral and decays exponentially with distance away from the edge. Different from the MZM bound to the $\pi$ flux in the above sections, the asympototic behavior for the edge Majorana modes has no power law decay factors in front of the exponential decay factor $e^{-|\Delta x/J|}$.

For the lattice Kitaev model Hamiltonian Eq.(\ref{eq:GS_Hamiltonian}) with an infinite  straight edge, however, the Hamiltonian has the form of Eq.(\ref{eq:bilinear_hamiltonian}) and from last section, the Majorna mode must have zero energy. The chiral edge modes with finite energy obtained from the continuum model are then not Majorana modes in the lattice model. Only the edge mode with $k_y=0$ is a Majorana mode in the lattice model which decays with distance to the edge as $\sim e^{-|\Delta x/J|}$.

In conclusions, we studied two types of defect induced Majorana zero mode in the ground state of the Kitaev model that decays exponentially in space and can be easily manipulated for braiding in quantum computing. In both approaches, we first apply a weak uniform $[111]$ magnetic field on the Kitaev model which turns the Kitaev model to an effective $p_x+ip_y$ superconductor of spinons. We then studied two specific defects, one is a vacancy in the Kitaev model, the other a polarized spin. Both defects correspond to a topologically trivial spot in the topologically non-trivial Kitaev system. We showed that both the single site vacancy and the fully polarized spin binds a vortex in the ground state at weak uniform $[111]$ magnetic field.   In both cases, the vortex results in a Majorana zero mode that decays exponentially in space and is robust against local non-magnetic perturbations and other Majorana zero modes far away. This is in contrast to the Majorana zero modes bound to pure gapless Kitaev model with vacancies. The Majorana zero modes we discussed in this work are then potential candidates for braiding in quantum computing.

Though the realization of the Kitaev model in real materials is still  challenging, it's possible to be realized in a cold-atom system~\cite{Duan2003}. At the same time, the fast development on single-site microscopy in optical lattices makes it promising to achieve and manipulate the Majorana zero mode we discussed in this work, especially through the locally polarized spin in the Kitaev model. In such case, a braiding procedure of Majorana zero modes may be performed according to the scheme discussed in a p-wave superconductor in Ref~\cite{Hoffman2019}.

\section{Acknowledgement}

This work is supported by the National NSF of China under Grant No. 11974166 and 11574134.

Note: After finishing writing this paper, we noticed Ref.\cite{Shanker2010, Vojta2016, Das2016} which discussed the effects of a local magnetic impurity coupling to the pure Kitaev model. These works mainly focus on the Kondo effect of the system and the impurity screened state of such system has some similarity with the system we study in this work but not exactly the same.

\section{Appendix}

In this appendix, we show the derivation of the equations the zero mode satisfies in the Kitaev model  with a vacancy, and the results that with $t=0$ and a vacancy on a single B site as shown in Fig.\ref{fig:lattice}a,  the amplitudes of the zero mode is nonzero only on the A sublattice for both the flux free and flux-threading case. Whereas for a vacancy on a single B site but with three local magnetic field $h^\alpha_\alpha$ on the neighboring site $\alpha=x, y, z$ as shown in Fig.\ref{fig:lattice}c, the amplitude of the zero mode locates only on the B sublattice and the three $\alpha$ sites at $t=0$.

We first consider the case with only a vacacncy site without local magnetic field. The lattice configuration is shown in Fig.\ref{fig:lattice}a in the main text and the vacancy is on a B sublattice site.
For simplicity, we label the Majorana fermion field $\hat{c}$ on the sublattice A and B as $\hat{a}_{l, j}$ and $\hat{b}_{l, j}$ respectively in this appendix.
In the flux free case, the Hamiltonian Eq.(\ref{eq:gap_Hamiltonian}) can then be written as
\begin{widetext}
\begin{eqnarray}
\hat{H}&=&-i\sum_{(l,j)}(J_x \hat{a}_{l,j}\hat{b}_{l+1,j-1}+J_y \hat{a}_{l,j} \hat{b}_{l+1,j}+J_z  \hat{a}_{l,j}\hat{b}_{l,j})\nonumber\\
&&+i\sum_{(l,j)}(J_x \hat{b}_{l,j}\hat{a}_{l-1,j+1}+J_y \hat{b}_{l,j}\hat{a}_{l-1,j}+J_z \hat{b}_{l,j}\hat{a}_{l,j}) \nonumber\\
&&+it\sum_{(l,j)}(-\hat{a}_{l,j}\hat{a}_{l,j-1}+\hat{a}_{l,j}\hat{a}_{l-1,j}-\hat{a}_{l,j}\hat{a}_{l-1,j+1}+\hat{a}_{l,j}\hat{a}_{l,j+1}-\hat{a}_{l,j}\hat{a}_{l+1,j}+\hat{a}_{l,j}\hat{a}_{l+1,j-1}) \nonumber\\
&&+it\sum_{(l,j)}(\hat{b}_{l,j}\hat{b}_{l,j-1}-\hat{b}_{l,j}\hat{b}_{l-1,j}+\hat{b}_{l,j}\hat{b}_{l-1,j+1}-\hat{b}_{l,j}\hat{b}_{l,j+1}+\hat{b}_{l,j}\hat{b}_{l+1,j}-\hat{b}_{l,j}\hat{b}_{l+1,j-1}). \nonumber\\
\end{eqnarray}
The zero mode satisfies $\hat{H}\hat{\psi}_0=0$, and $\hat{\psi}_0$ may be expressed as $\hat{\psi}_0=\sum_{l,j} a_{l,j}\hat{a}_{l,j} +b_{l,j} \hat{b}_{l,j}$,  where $a_{l,j}$ and $b_{l,j}$ is the coefficient of the field $\hat{a}_{l,j}$ and $\hat{b}_{l,j}$ on the A and B sublattice respectively. From the Hamiltonian matrix in the above, it's then easy to obtain the equations  the zero mode satisfies:
\begin{eqnarray}
&&(J_x a_{l-1,j+1}+J_y a_{l-1,j}+J_z a_{l,j})
 -t[(b_{l,j+1}-b_{l,j-1})+(b_{l-1,j}-b_{l-1,j+1})+(b_{l+1,j-1}-b_{l+1,j})]=0  \label{eq:A_and_B_app}\\
&&(J_x b_{l+1,j-1}+J_y b_{l+1,j}+J_z b_{l,j})
 -t[(a_{l,j+1}-a_{l,j-1})+(a_{l-1,j}-a_{l-1,j+1})+(a_{l+1,j-1}-a_{l+1,j})]=0  \label{eq:B_and_A_app}
\end{eqnarray}
These equations hold true everywhere except near the vacancy  with $l,j=-1, 0, 1$. Near the vacancy, the zero mode satisfies the boundary condition in the main text, i.e., (a) Eq.(\ref{eq:A_and_B_app}) does not exist at $(l,j)=(0, 0)$; (b)For other $l, j=-1, 0, 1$, Eq.(\ref{eq:A_and_B_app})  and Eq.(\ref{eq:B_and_A_app}) is satisfied with the amplitude $b_{0, 0}=0$.

\subsection{At $t=0$ }
The above zero mode is not solvable analytically for finite $t$. But at $t=0$, the zero mode may be solved by dividing the lattice to two parts $l> 1$ and $l<-1$, and match the boundary condition at $l=0,1,-1$. This zero mode is solved in Ref~\cite{Pereira2006, Santhosh2012} at the flux free case. We recapitulate the process here briefly and show some of the detailed derivation that's absent in the reference but also needed to study the case with local magnetic fields. For simplicity, we only consider the isotropic gapless case $J_x=J_y=J_z=J$ here.

At $t=0$, Eq.(\ref{eq:A_and_B_app})  and Eq.(\ref{eq:B_and_A_app}) may be solved separately in the $l>0$ adn $l<-1$ regions. In these regions, the system has translation symmetry in the $j$ direction. 
By a Fourier transformation $a_{l,j}=\sum_{k}a_{l}(k) e^{ik j},b_{l,j}=\sum_{k}b_{l}(k) e^{ikj},k=\frac{2\pi m}{N},m=1,2...,N$, one gets
\begin{eqnarray}
a_l(k)&=&[-(1+e^{ik})]^l a_0(k), \nonumber\\
b_l(k)&=&[-(1+e^{-ik})]^{-(l-1)}b_1(k), \label{eq:recursion_A}\nonumber\\
\end{eqnarray}
for $l>0$ and 
\begin{eqnarray}
a_l(k)&=&[-(1+e^{ik})]^{l+1}a_{-1}(k), \nonumber\\
b_l(k)&=&[-(1+e^{-ik})]^{-(l+1)}b_{-1}(k), \label{eq:recursion_B}
\nonumber\\
\end{eqnarray}
for $l<-1$.

To have a decay solution, the recursion factors of both $a_l(k)$ and $b_l(k)$ need to have modulus less than one. From the above recursion relationships, one can see that the decay range of $k$ for the coefficients $a_l(k)$ (or $b_l(k)$) at $l>0$ and $l<-1$ is complementary. And the decay range of $a_l(k)$ corresponds to the increase range of $b_l(k)$ so $a_l(k)$ and $b_l(k)$ cannot both be nonzero for a given $k$.

Here $a_0(k), a_{-1}(k), b_1(k), b_{-1}(k)$ are determined by matching the boundary condition. In the flux free case, the boundary condition for $a_{l, j}$ is 
$a_{-1, j}+a_{-1, j+1}+a_{0,j}=0$ except for $j=0$, i.e.
\begin{equation}
\sum_k e^{ikj}[a_0(k)+(1+e^{ik})a_{-1}(k)]=0, \ j\neq 0.
\end{equation}
This results in $a_0(k)+(1+e^{ik})a_{-1}(k)={\rm const}$ since the Fourier transformation of a $\delta$ function is a constant. From the above analysis, for a given $k$, only one of $a_0(k)$ and $a_{-1}(k)$ may be nonzero.

The boundary condition for $b_{l,j}$ in the flux free case is 
\begin{eqnarray}
&& b_{1, j}+b_{1, j-1}+b_{0, j}=0, \ j\neq 0,  \nonumber\\
&& b_{0, j}+b_{0, j+1}+b_{-1, j+1}=0,\  j\neq 0, -1, \nonumber\\
&& b_{1, 0}+b_{1,-1}=0, \ b_{0, 1}+b_{-1, 1}=0, \ b_{0, -1}+b_{-1, 0}=0.
\end{eqnarray}
These correspond to the equations for $b_l(k)$ as
\begin{eqnarray}
&&\sum_k e^{ikj}[b_0(k)+(1+e^{-ik})b_{1}(k)]=0, \ j\neq 0,\\
&&\sum_k e^{ikj}[b_{-1}(k)e^{ik}+(1+e^{ik})b_{0}(k)]=0, \ j\neq 0, -1 \\
&&\sum_k (1+e^{-ik})b_{1}(k)=0, \ \sum_k e^{ik} [b_0(k)+b_{-1}(k)]=0, \ \sum_k [e^{-ik}b_0(k)+b_{-1}(k)]=0.
\end{eqnarray}
This results in 
\begin{eqnarray}
&&b_0(k)+(1+e^{-ik})b_{1}(k)={\rm const}, \label{eq:bc_B}\\
&&b_{-1}(k)+(1+e^{-ik})b_{0}(k)={\rm const},  \\
&&\sum_k (1+e^{-ik})b_{1}(k)=0, \ \sum_k e^{ik} [b_0(k)+b_{-1}(k)]=0, \ \sum_k [e^{-ik}b_0(k)+b_{-1}(k)]=0.
\end{eqnarray}
These equations,  together with the condition that  $b_{-1}(k)$ and $b_1(k)$ cannot both be nonzero for any given $k$, result in $b_0(k)=b_1(k)=b_{-1}(k)=0$ for all $k$ which leads to $b_l(k)=0$ for all $l$, i.e., the coefficients of the zero mode on the B sublattice must be zero at $t=0$.

The amplitude of the zero mode is then nonzero only on the A sublattice, i.e., the sublattice opposite to that of the vacancy site. One can then choose $a_0(k)=\Theta[1-|f(k)|]$ and $(1+e^{ik}) a_{-1}(k)=\Theta[|f(k)|-1]$, where $f(k)\equiv 1+e^{ik}$. The zero mode in this case may then be solved from the recursion relationship of $a_l(k)$ and the initial condition of $a_0(k)$ and $a_{-1}(k)$ in the above, and then  inverse Fourier transformation to the real space, as shown in Ref~\cite{Pereira2006, Santhosh2012}.

\

{\it We next consider the case with a flux threading the vacancy. }

In this case, the recursion relationship for $a_l(k)$ and $b_l(k)$ in Eq.(\ref{eq:recursion_A}) and Eq.(\ref{eq:recursion_B}) still holds. However, the boundary condition at $l=-1, 0, 1$ becomes
\begin{eqnarray}
&& a_{-1, j}+a_{-1, j+1}-\sign[j] a_{0,j}=0, \ j \neq 0,  \nonumber\\
&& b_{1, j}+b_{1, j-1}-\sign[j] b_{0, j}=0, \ j \neq 0,  \nonumber\\
&& b_{0, j}+b_{0, j+1}+b_{-1, j+1}=0,\  j \neq 0, -1, \nonumber\\
&& b_{1, 0}+b_{1,-1}=0, \ b_{0, 1}+b_{-1, 1}=0, \ b_{0, -1}+b_{-1, 0}=0.
\end{eqnarray}

Since the recursion relationship in Eq.(\ref{eq:recursion_A}) and Eq.(\ref{eq:recursion_B}) still hold, the constraint that $b_{-1}(k)$ and $b_1(k)$ cannot  both be nonzero for a given $k$ still works. Yet due to the flux, the boundary condition Eq.(\ref{eq:bc_B}) for the B sublattice is no longer true. However, it's easy to see that the solution $b_l(k)=0$ and so $b_{ij}=0$ still hold for the B sublattice. The amplitude of the zero mode then still locates only on the A sublattice in this case.

\

{\it At last, we  consider the case with a flux threading the vacancy and three local  magnetic field $h^\alpha_\alpha$ acting on the three neighboring $\alpha$ sites as shown in Fig.\ref{fig:lattice}c. }

In this case, there are three local magnetic field acting on the A sites of unit cell $(0,0), (-1, 0), (-1, 1)$. The Majorana fields $\hat{c}$ on these three sites then couple not only to the $\hat{c}$ field on the neighboring B sites, but also the $\hat{b}^\alpha_\alpha$ Majorana fermion on the same site. The recursion relationship at $l>0$ and $l<-1$ still holds.  But the boundary condition the zero mode satisfies near the vacancy then becomes
\begin{eqnarray}
&& a_{-1, j}+a_{-1, j+1}-\sign[j] a_{0,j}=0, \ j \neq 0,  \nonumber\\
&& b_{1, j}+b_{1, j-1}-\sign[j] b_{0, j}=0, \ j \neq 0,  \nonumber\\
&& b_{0, j}+b_{0, j+1}+b_{-1, j+1}=0,\  j \neq 0, -1, \nonumber\\
&& b_{1, 0}+b_{1,-1}+\langle \sigma^z_0\rangle b^z_{0,0}=0, \ b_{0, 1}+b_{-1, 1}+\langle \sigma^x_0\rangle b^x_{-1,1}=0, \ b_{0, -1}+b_{-1, 0}+\langle \sigma^y_0\rangle b^y_{-1, 0}=0.\label{eq:coupling_local_field}
\end{eqnarray}
Here $\langle \sigma^\alpha_0 \rangle=1/\sqrt{3}, \alpha=x, y, z$ is the magnetization of the spin on the site $0$, $b^{\alpha}_{lj}$ is the amplitude of the $\hat{b}^\alpha_\alpha$ Majorana field on the A sublattice of unit cell $(l, j)$. From Eq.(\ref{eq:coupling_local_field}), we see that due to the  local magnetic field on the three $\alpha$ sites, the amplitudes $b_{-1, j}, b_{1, j}$ on the B sublattice are no longer zero since the local magnetic field results in finite $b_{lj}^\alpha$ on the   three $\alpha$ sites. This results in a finite $b_{-1}(k)$ and finite $b_1(k)$. Since the decay range of $k$ for $a_l(k)$ and $b_l(k)$ is complementary, a finite $b_l(k)$ results in zero of $a_l(k)$. For the reason, the amplitude $a_{ij}$ of the zero mode on the A sublattice in this case is zero, as obtained in our numerics. Note that the amplitudes $b^z_{0,0}, b^x_{-1, 1}, b^y_{-1, 0}$ we obtained on the three A sites neighboring to site $0$ are finite, but they are amplitudes of the bond Majorana fermions, not the matter majorana fermions $\hat{b}_{lj}$.

\end{widetext}


\begin{thebibliography}{99}


\bibitem{Read2000}Read N and Green D 2000 {\it Phys. Rev. B} \textbf{61} 10267


\bibitem{Kitaev2001} Kitaev A.Yu  2001 {\it Phys-Usp.} \textbf{44}  131


\bibitem{Kouwenhoven2012}Mourik V,  Zuo K,  Frolov S M, Plissard S R, Bakkers E P A M, L.P. Kouwenhoven  2012 {\it Science} \textbf{336} 1003


\bibitem{Yazdani2017}Jeon S, Xie Y,  Li J, Wang Z,  Bernevig B A, Yazdani A  2017 {\it Science} \textbf{358} 772


\bibitem{Kouwenhoven2019}Zhang H, Liu D E, Wimmer M, and Kouwenhoven L P 2019 {\it Nat. Commun.} \textbf{10} 5128

\bibitem{Oreg2010} Oreg Y, Refael G, and von Oppen F, 2010 {\it Phys. Rev. Lett.} \textbf{105} 177002


\bibitem{Moore1991}Moore G and Read N 1991 {\it Nucl. Phys. B} \textbf{360} 362



\bibitem{Silaev2010}Silaev M A and Volovik G E 2010 {\it J. of Low Temp. Phys.} \textbf{161}  460


\bibitem{Sau2010}Sau J D,  Lutchyn R M, Tewari S and Das Sarma S 2010 {\it Phys. Rev. Lett.} \textbf{104} 040502


\bibitem{Sau2012}Sau J D and  Das Sarma S, 2012 {\it Nat. Commun.} \textbf{3} 964


\bibitem{Fulga2013}Choy T P, Edge J M,  Akhmerov A R, and Beenakker C W J 2011 {\it Phys. Rev. B} \textbf{84} 195442




\bibitem{Yazdani2013}Nadj-Perge S, Drozdov I K,  Bernevig B A, and Yazdani A 2013 {\it Phys. Rev. B} \textbf{88} 020407(R)


\bibitem{Rice1995}Rice T M  and Sigrist M 1995 {\it J. Phys: Cond. Matt.} \textbf{7} L643 

\bibitem{Fu2008}Fu L and Kane C L 2008 {\it Phys. Rev. Lett.} \textbf{100} 096407








\bibitem{Nayak2008}Nayak C, Simon S H, Stern A, Freedman M and  Das Sarma S 2008 {\it Rev. Mod. Phys.} \textbf{80} 1083 


\bibitem{Kitaev2006}Kitaev A 2006 {\it Ann. of Phys.} \textbf{321} 2

\bibitem{Xiang2007-2}Feng X S, Zhang G M, and Xiang T, 2007 {\it Phys. Rev. Lett.} {\textbf 98} 087204


\bibitem{Motome2019}Motome Y and Nasu J 2020  {\it J  Phys. Soc.  Jpn.} \textbf{89} 012002



\bibitem{Willans2010}Willans A J, Chalker J T, and Moessner R 2010 {\it Phys. Rev. Lett.} \textbf{104} 237203

\bibitem{Willans2011}Willans A J, Chalker J T, and Moessner R 2011  {\it Phys. Rev. B} \textbf{84} 115146








\bibitem{Volovik1999}Volovik G E 1999 {\it JETP Lett.} \textbf{70} 609


\bibitem{Xiang2007} Lee D H, Zhang G M, and  Xiang T 2007 {\it Phys. Rev. Lett.} \textbf{99} 196805



\bibitem{Ivanov2001}Ivanov D A 2001 {\it Phys. Rev. Lett.} \textbf{86} 268




\bibitem{Santhosh2012} Santhosh G, Sreenath V,  Lakshminarayan A, and Narayanan R 2012 {\it Phys. Rev. B} \textbf{85} 054204




\bibitem{Pereira2006} Pereira V M, Guinea F,  Lopes dos Santos J M B, Peres  N M  R  and Castro Neto A H 2006 {\it Phys. Rev. Lett.} \textbf{96}  036801



\bibitem{Jiang2020}Jiang M H,  Liang S, Chen W, Qi Y, Li J X, and Wang Q H, 2020 {\it Phys. Rev. Lett.} \textbf{125}, 177203

\bibitem{Liang2018} Liang S, He BS,  Dong Z Y, Chen W, Li J X and Wang Q H 2018 {\it Phys. Rev. B } \textbf{98} 104410

\bibitem{Liang2018_2}Liang S, Jiang  M H, Chen W,  Li J X and Wang Q H 2018 {\it Phys. Rev. B} \textbf{98} 054433




\bibitem{Duan2003}Duan L M,  Demler E, and Lukin M D 2003 {\it Phys. Rev. Lett.} \textbf{91} 090402


\bibitem{Hoffman2019}November B H, Sau J D, Williams J R, and Hoffman J E 2019 arXiv:1905.09792[cond-mat]


 \bibitem{Note}The magnetization curve we obtained shows hysteresis when the local magnetic field is swept in the increase and decrease direction respectively. This also indicates that the phase transition at the polarization is first order.
 
 
 \bibitem{Otten2019} Otten D,  Roy A, and Hassler F 2019 {\it Phys. Rev. B.} {\bf 99} 035137
 
\bibitem{Shanker2010}Dhochak K, Shankar R and Tripathi V 2010  {\it Phys. Rev. Lett.} {\bf 105} 117201

\bibitem{Vojta2016}Vojta M, Mitchell A K  and Zschocke F 2016 {\it Phys. Rev. Lett.} {\bf 117} 037202

\bibitem{Das2016} Das S D, Dhochak K and Tripathi V 2016 {\it Phys. Rev. B} {\bf 94} 024411



\end{thebibliography}
\end{document}